\documentclass[sigconf]{acmart}
\renewcommand\footnotetextcopyrightpermission[1]{} %remove copyright
\usepackage{graphicx}
\usepackage{booktabs}
\usepackage{bbding}
\usepackage{enumitem}
\usepackage{multirow}
\usepackage{verbatim}
\usepackage{subfiles}
\usepackage{xcolor}
\usepackage{hyperref}
\usepackage[table]{xcolor}
\usepackage[most]{tcolorbox}
\usepackage[utf8]{inputenc}
\newtcolorbox{PromptBox}[2][]{
    enhanced,
    breakable,
    colback=gray!5,
    colframe=gray!80,
    fonttitle=\bfseries,
    title=#2,
    width=\columnwidth,
    sharp corners=south,
    boxrule=0.5pt,
    top=2mm,
    bottom=2mm,
    before skip=10pt,
    after skip=10pt,
    #1
}
\newtcolorbox{TraceBox}[2][]{
    enhanced,
    breakable,
    colback=blue!5!white,
    colframe=blue!50!black,
    fonttitle=\bfseries,
    title=#2,
    width=\columnwidth,
    sharp corners=south,
    boxrule=0.5pt,
    top=2mm,
    bottom=2mm,
    before skip=10pt,
    after skip=10pt,
    #1
}
\usepackage{listings}

\usepackage[dvipsnames]{xcolor}
\NewDocumentCommand{\lifu}{mO{}}{%
  \textcolor{red}{\textsuperscript{\textit{Lifu}}\textsf{\textbf{\small[#1]}}}%
}
\NewDocumentCommand{\zihao}{mO{}}{%
  \textcolor{Orange}{\textsuperscript{\textit{zihao}}\textsf{\textbf{\small[#1]}}}%
}
\NewDocumentCommand{\muyang}{mO{}}{%
  \textcolor{Blue}{\textsuperscript{\textit{muyang}}\textsf{\textbf{\small[#1]}}}%
}
\NewDocumentCommand{\haibo}{mO{}}{%
  \textcolor{Green}{\textsuperscript{\textit{haibo}}\textsf{\textbf{\small[#1]}}}%
}

\newcommand{\dataname}{\textsc{VideoGlitchBench}}
\newcommand{\methodname}{\textsc{GliDe}}

\AtBeginDocument{%
  }

\setcopyright{acmlicensed}
\copyrightyear{2026}
\acmYear{2026}
\acmDOI{XXXXXXX.XXXXXXX}
\acmConference[ACM Multimedia 2026]{Make sure to enter the correct conference title from your rights confirmation email}{Rio de Janeiro}{Brazil}
\acmISBN{978-1-4503-XXXX-X/2018/06}

\acmSubmissionID{6212}

\begin{document}

\title{Open-Ended Video Game Glitch Detection with Agentic Reasoning and Temporal Grounding}

% Author information in ACM format
\author{Muyang Zheng}
\affiliation{%
  \institution{University of California, Davis}
  \city{Davis, California}
  \country{USA}
}
\email{muyzheng@ucdavis.edu}

\author{Tong Zhou}
\affiliation{%
  \institution{Virginia Polytechnic Institute and State University}
  \city{Blacksburg, Virginia}
  \country{USA}
}
%\email{tongzhou@vt.edu}

\author{Geyang Wu}
\affiliation{%
  \institution{University of California, Davis}
  \city{Davis, California}
  \country{USA}
}
%\email{gywu@ucdavis.edu}

\author{Zihao Lin}
\affiliation{%
  \institution{University of California, Davis}
  \city{Davis, California}
  \country{USA}
}
%\email{qzlin@ucdavis.edu}

\author{Haibo Wang}
\affiliation{%
  \institution{University of California, Davis}
  \city{Davis, California}
  \country{USA}
}
%\email{hibwang@ucdavis.edu}

\author{Lifu Huang}
\affiliation{%
  \institution{University of California, Davis}
  \city{Davis, California}
  \country{USA}
}
\email{lfuhuang@ucdavis.edu}

% By default, the full list of authors will be used in the page headers.
%\renewcommand{\shortauthors}{First Author and Second Author}

\begin{abstract}
\textit{Open-ended video game glitch detection} aims to identify glitches in gameplay videos, describe them in natural language, and localize when they occur. Unlike conventional game glitch understanding tasks which have largely been framed as image-level recognition or closed-form question answering, this task requires reasoning about game-specific dynamics such as mechanics, physics, rendering, animation, and expected state transitions directly over continuous gameplay videos and distinguishing true glitches from unusual but valid in-game events. 
To support this task, we introduce \textbf{\dataname}, the first benchmark for open-ended video game glitch detection with temporal localization. \dataname{} contains 5,238 gameplay videos from 120 games, each annotated with detailed glitch descriptions and precise temporal spans, enabling unified evaluation of semantic understanding and temporal grounding.
We further propose \textbf{\methodname}, an agentic framework with three key components: a game-aware contextual memory for informed reasoning, a debate-based reflector for multi-perspective glitch detection and verification, and an event-level grounding module that recovers complete glitch intervals from fragmented temporal evidence. We also design a task-specific evaluation protocol that jointly measures semantic fidelity and temporal accuracy. Experiments show that this task remains highly challenging for current multimodal models, while \methodname{} achieves substantially stronger performance than corresponding vanilla model baselines.
Project Page: \href{https://sandyyyzheng.github.io/glide-project-page/}{\textcolor{blue}{sandyyyzheng.github.io/glide-project-page}}

\end{abstract}

%%
%% The code below is generated by the tool at http://dl.acm.org/ccs.cfm.
%% Please copy and paste the code instead of the example below.
%%


%%
%% Keywords. The author(s) should pick words that accurately describe
%% the work being presented. Separate the keywords with commas.
%% A "teaser" image appears between the author and affiliation
%% information and the body of the document, and typically spans the
%% page.

% \received{20 February 2007}
% \received[revised]{12 March 2009}
% \received[accepted]{5 June 2009}

%%
%% This command processes the author and affiliation and title
%% information and builds the first part of the formatted document.
\begin{teaserfigure}
    \centering
    \includegraphics[width=\textwidth]{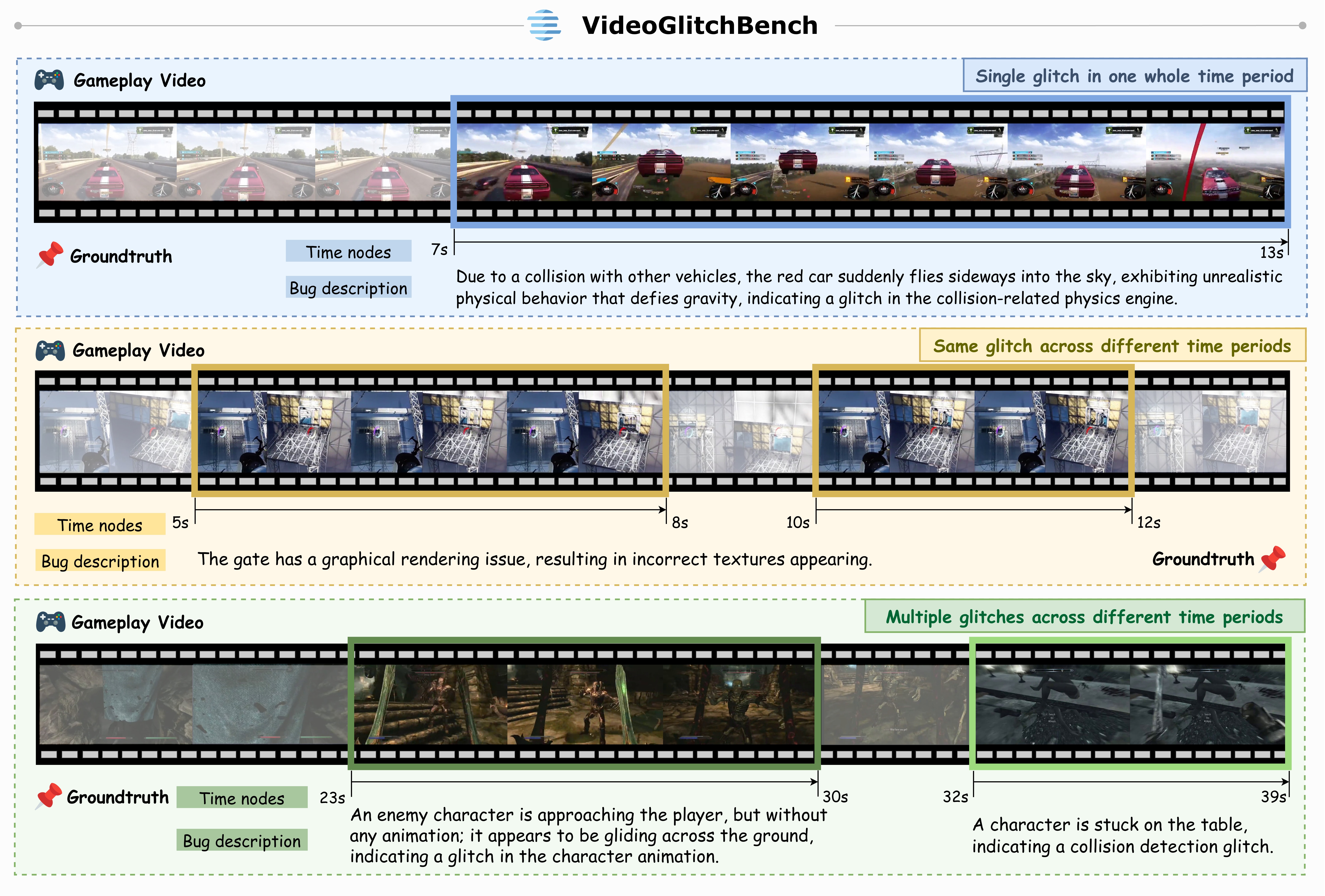}
    \caption{We introduce \dataname, a comprehensive benchmark designed for open-ended video game glitch detection.}
    \Description{Examples of our benchmark \dataname.\lifu{to save space, you can probably make this figure denser, and use 1-2 example videos to demonstrate these three cases; also add the category of the bugs to highlight the requirement of diverse types of reasoning for mechanics, physics, rendering, etc.}} 
    \label{fig:dataset}
\end{teaserfigure}

\maketitle

\begin{table*}[ht]
\centering
\small
\caption{Comparison between \dataname{} and existing game glitch datasets.
% \lifu{you can move this to appendix if more space is needed}
%\zihao{Samples means "videos" instead of "games" right? Need to make sure the description is consistent with your writing paragraphs.} \muyang{Yes. I wrote 'samples' because some datasets are images instead of videos. I have modified 'samples' to 'scale' and clarified the modality.} \zihao{You can add another column to present the difference of the evaluation metrics between your work and previous work.} \muyang{Done.}
}
\label{tab:dataset_comparison}
\begin{tabular}{@{}llllcl@{}}
\toprule
\textbf{Dataset} & \textbf{Scale} & \textbf{Pipeline} & \textbf{Annotation} & \textbf{Localization} & \textbf{Evaluation} \\ \midrule
GamePhysics~\cite{clipxgamephysics} & 26,954 videos & Semi-Auto & Game name/weak metadata & \XSolidBrush &  Retrieval relevance\\
GameBugDescriptions~\cite{gamebugdesc} & 167 videos & Manual & Event description \& Bug type & \XSolidBrush & Multiple-choice accuracy\\
GlitchBench~\cite{glitchbench} & 923 images & Manual & Short glitch description & \XSolidBrush & Description quality \\
VideoGameBunny~\cite{videogamebunny} & 185,259 images & Semi-Auto & \begin{tabular}[t]{@{}l@{}}Caption \& Image-to-JSON \\ \& Question-answer pairs\end{tabular} & \XSolidBrush & N/A (Designed for instruction-tuning) \\
GameBench~\cite{physgame} & 880 videos & Semi-Auto & Multiple-choice questions & \XSolidBrush & Multiple-choice accuracy\\
PhysGame~\cite{physgame} & 38,957 videos & Semi-Auto & Question-answer pairs & \XSolidBrush & N/A (Designed for instruction-tuning) \\
\textbf{\dataname} & \textbf{5,238 videos} & \textbf{Semi-Auto} & \begin{tabular}[t]{@{}l@{}} \textbf{Detailed glitch description} \\ \textbf{\& start/end timestamps} \end{tabular} & \CheckmarkBold & \begin{tabular}[t]{@{}l@{}} \textbf{Description quality \& grounding} \\ \textbf{accuracy} \end{tabular} \\ \bottomrule
\end{tabular}
\end{table*}

\section{Introduction}

Game glitches are unintended failures in gameplay, such as broken physics, rendering artifacts, animation errors, collision failures, or logic inconsistencies~\cite{lin2019identifybugs, wilkins20223dbugs}. 
Detecting such glitches from gameplay videos is an important capability for game quality assurance (QA)~\cite{taesirivideogameqa}, where testers typically examine gameplay sessions to inspect suspicious behaviors and summarize the issue with detailed bug reports. 
%need to determine whether a true glitch occurs, explain what went wrong, and identify when the event begins and ends. 
Recent works have begun to explore game glitch understanding through tasks such as image-level glitch detection~\cite{videogamebunny, glitchbench}, text-only bug reasoning~\cite{gamebugdesc}, and closed-form (e.g., multiple-choice or yes/no) question answering over gameplay videos~\cite{physgame, physvlm}. 
While these settings provide testbeds for studying glitch understanding, they address only restricted aspects of the problem and fall short of the richer contextual analysis in real-world gameplay-based QA.

To close this gap, we propose \textit{open-ended video game glitch detection}, a task in which a model must reason over raw gameplay videos, identify glitches in an open-ended manner, describe them in natural language, and temporally localize their occurrence as evidence for follow-up analysis. 
Compared with prior game glitch understanding tasks, this setting requires joint video understanding, game-aware reasoning, and temporal grounding in a unified framework. 
To support research on this problem, we introduce \textbf{\dataname}, the first benchmark for open-ended video game glitch detection with temporal localization. 
\dataname{} is constructed from community-reported gameplay videos in GamePhysics~\cite{clipxgamephysics} through careful game category-based selection, filtering, and a semi-automated annotation pipeline, where GPT-4o~\cite{hurst2024gpt4o} generates initial glitch descriptions from short video segments with Reddit discussion context, followed by human review and manual start--end timestamp annotation, yielding 5,238  high-quality, temporally grounded glitch descriptions for 120 games spanning diverse genres and glitch phenomena.

Distinct from previous benchmarks for video understanding or video anomaly detection~\cite{sultani2018vadsurvey, ramachandra2020vadsurvey, luo2017shanghaitech, sultani2018ucfcrime, wu2020xdviolence}, which mainly focus on identifying specific events in real-world scenes such as surveillance, traffic, or industrial monitoring, \dataname{} reveals two unique and critical challenges. \textbf{First}, a model must distinguish genuine glitches from behaviors that may appear abnormal visually but are actually consistent with the game’s design, mechanics, and world logic. This requires understanding not only the visual scene itself, but also the behaviors of characters and objects, the ongoing gameplay context, and the broader game background. For example, an object such as ``\textit{a boat flying through the sky}'' may indicate a physics failure in one game, but could be a perfectly valid event in another game with stylized or fantastical mechanics. \textbf{Second}, glitches exhibit diverse temporal patterns: some occur only briefly, while others persist over extended periods, and the same underlying glitch may reappear multiple times in a video with gaps in between. A model must therefore not only detect local failure evidence, but also determine when temporally separated observations should be linked to the same glitch event. Figure~\ref{fig:dataset} shows typical examples in \dataname, including single glitch, repeated occurrences of the same glitch, and multiple glitches within one video.

To address these challenges, we propose \textbf{\methodname}, an agentic framework for open-ended video \underline{G}ame g\underline{LI}tch \underline{DE}tection with temporal localization. The design of \methodname{} is directly motivated by the structure of the task. To determine whether a suspicious event is truly a glitch in a particular game context, \methodname{} first builds a \textit{game-aware contextual memory} that incrementally accumulates cues about the scene, activities, interactions, and local gameplay dynamics, thereby providing a stronger prior for later judgments. To reduce false positives caused by visually unusual yet valid gameplay behaviors, \methodname{} then performs \textit{multi-step verification} through adaptive tool use and a debate-based reflector, encouraging the model to compare a glitch hypothesis against plausible in-game explanations before reaching a decision. Finally, to handle glitches that are short-lived, long-lasting, or intermittently recurring, \methodname{} includes an \textit{event-level grounding} module that consolidates fragmented evidence across windows and recovers complete glitch intervals, including multiple disjoint occurrences of the same glitch. 

We also design a new evaluation protocol to match and compare a set of system-generated, temporally grounded glitch descriptions against the ground-truth annotations for each video, by
jointly assessing their semantic matching through LLM-based scoring and temporal grounding using temporal IoU. Extensive experiments show that \dataname{} is highly challenging for current video understanding models. Even strong proprietary models such as gemini-2.0-flash~\cite{gemini2} and claude-3.5-haiku~\cite{claude3.5} achieve limited performance, with the best baseline reaching only 26.01\% F1 and 0.47 mIoU, while the best vanilla open-source model achieves 21.62\% F1 and the average F1 across six open-source backbones is only 14.47\%, highlighting the difficulty of jointly producing accurate open-ended glitch descriptions and precise temporal localization. In contrast, \methodname{} consistently improves both semantic understanding and grounding across all evaluated backbones: averaged over six open-source models, it raises F1 from 14.47\% to 36.05\% and mIoU from 0.28 to 0.51, and all \methodname{}-enhanced models outperform the reported proprietary baselines on the overall metrics. These results demonstrate both the difficulty of \dataname{} and the effectiveness of \methodname{} as a unified solution for context-aware, temporally grounded glitch detection. 

Our main contributions are summarized as follows: 
\textbf{(1)} We introduce \dataname{}, the first benchmark for open-ended video game glitch detection with temporal localization, featuring 5,238 gameplay videos across 120 games with detailed descriptions and precise timestamps. 
\textbf{(2)} We propose \methodname{}, an agentic framework that tackles glitch detection challenges through a game-aware contextual memory, a debate-based verification mechanism, and an event-level grounding module to consolidate fragmented glitch intervals. 
\textbf{(3)} We design a comprehensive evaluation protocol that jointly assesses semantic fidelity and temporal localization. Extensive experiments demonstrate that \methodname{} significantly outperforms competitive baselines on this challenging benchmark.

\section{Related Work}

%\subsection{Video Game Glitch Detection}

%\par \textbf{Game Glitch Benchmarks.} 

\textbf{Game Glitch Understanding Benchmarks. } %Benchmark construction has played a central role in advancing game glitch understanding. %Establishing robust datasets and evaluation metrics has been a primary focus to quantify model capabilities in this domain. 
GamePhysics~\cite{clipxgamephysics} introduced the first large-scale dataset of gameplay videos, containing more than 26,000 videos from over 1,800 games. Building on this, more recent research has shifted toward constructing specialized benchmarks for evaluating multimodal large language models (MLLMs) on glitch understanding tasks. 
GlitchBench~\cite{glitchbench} evaluates MLLMs on recognizing unusual gameplay scenarios, focusing on single-frame visual glitch detection. GameBugDescriptions~\cite{gamebugdesc} tests zero-shot bug detection via multiple-choice questions based on text descriptions, while GameBench~\cite{physgame} uses expert-annotated gameplay videos paired with multiple-choice questions to assess visual physical reasoning capabilities. To train these capabilities, PhysGame~\cite{physgame} provides a large-scale video instruction dataset of over 140,000 QA pairs designed to enhance physical world understanding.
% GlitchBench~\cite{glitchbench} evaluates MLLMs on recognizing unusual gameplay scenarios, focusing on single-frame visual glitch detection. For text-based evaluation, GameBugDescriptions~\cite{gamebugdesc} relies solely on step-by-step textual descriptions of gameplay events, testing bug detection in language models via multiple-choice questions. Shifting to multimodal evaluation, GameBench~\cite{physgame} uses expert-annotated gameplay videos paired with multiple-choice questions to assess visual physical reasoning capabilities. Finally, to actively train these capabilities rather than just evaluate them, PhysGame~\cite{physgame} provides a large-scale, video-based instruction dataset of over 140,000 question-answer pairs designed to enhance physical world understanding.
% However, these benchmarks either focus on image-level detection or question-answering formats, and none of them provide temporal localization. As a result, they do not fully reflect the practical setting of game QA, where a model must detect glitches from raw videos, describe them in an open-ended manner, and identify when they occur. In contrast, our \dataname{} targets open-ended video game glitch detection with temporal localization, providing both natural-language glitch descriptions and precise timestamps. Table~\ref{tab:dataset_comparison} summarizes the key differences between \dataname{}h and existing game glitch datasets.
However, restricted to image-level detection or multiple-choice QA formats without temporal localization, these benchmarks fail to reflect the practical demands of game QA. Addressing this gap, \dataname{} targets open-ended glitch detection directly from raw videos, providing both natural language descriptions and precise timestamps. Table~\ref{tab:dataset_comparison} summarizes these key differences.

% \zihao{Need to add one or two sentences to describe the limitations of previous benchmarks, and claim the differences (novelty) of your work compared to previous benchmarks.} \muyang{Done.}

\textbf{Game Glitch Detection Methods.} Existing methods for gameplay glitch detection have largely been shaped by the scope of previous benchmarks, and thus mainly operate through retrieval, question answering, or direct single-pass prediction. For example,  GamePhysics~\cite{clipxgamephysics} employs a CLIP-based retrieval approach to identify bug-related gameplay clips from textual queries, demonstrating that pretrained vision–language models can capture useful signals for glitch-related events. VideoGameBunny~\cite{videogamebunny} introduces a game-oriented multimodal assistant trained on large-scale gameplay instruction data, while PhysVLM~\cite{physvlm} enhances video language models with physical commonsense supervision to better detect violations of physics rules in gameplay videos. They are effective for matching predefined queries or selecting from limited answer spaces, but are insufficient for open-ended glitch detection, where models must reason over and identify genuine glitches directly from raw videos with precise temporal localization.

\textbf{Agentic Frameworks for Video Understanding.} 
For complex tasks requiring selective search and temporal grounding, agentic pipelines are increasingly replacing single-pass predictions~\cite{tang2025videosurvey}. Systems like VideoAgent~\cite{wang2024videoagent}, TraveLER~\cite{shang2024traveler}, and VideoMind~\cite{liu2025videomind} successfully leverage iterative planning, modular tools, and evidence collection to enhance video QA. Concurrently, in the gaming domain, automated testing agents~\cite{autoplaysurvey} utilizing reinforcement learning~\cite{bergdahl2020drlgametest, ariyurek2019gametestagents} and exploration systems (e.g., CCPT~\cite{sestini2022ccpt}, Inspector~\cite{liu2022inspector}) are widely deployed to uncover bugs. However, these game agents focus strictly on active, interactive environment exploration, leaving a crucial gap for agentic frameworks designed to detect and temporally localize glitches from recorded gameplay videos.

\begin{figure*}[ht]
\centering
\includegraphics[width=\textwidth]{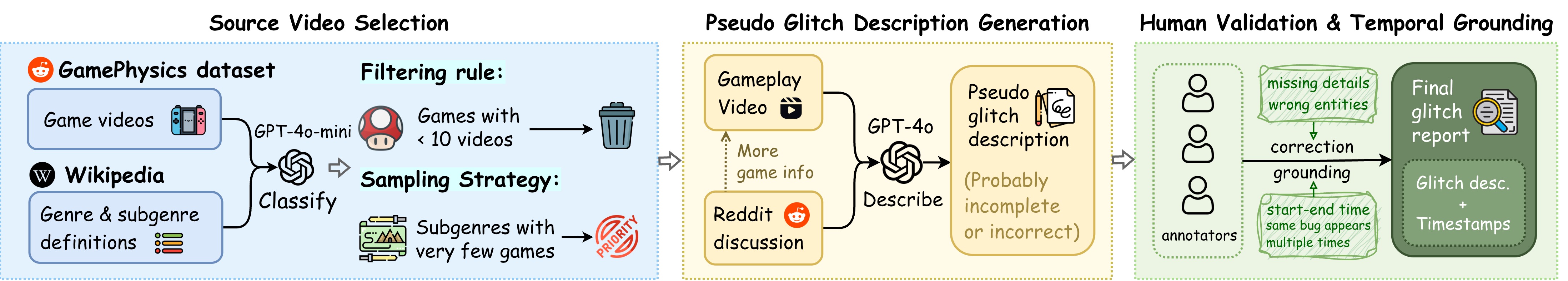}
\caption{The annotation pipeline of our benchmark \dataname.} \label{fig:annotation}
\Description{Benchmark curation.}
\end{figure*}

\textbf{Video Anomaly Detection (VAD)} aims to identify events deviating from normal patterns, traditionally focusing on intelligent surveillance~\cite{sultani2018vadsurvey}. While early methods relied on deep neural networks to model normality and detect deviations~\cite{hasan2016dnnvad, sultani2018vadsurvey, liu2018dnnvad, gong2019dnnvad}, recent MLLM-based approaches have shifted towards semantic reasoning over abnormal events~\cite{ramachandra2020vadsurvey}. Representative methods such as Holmes-VAD~\cite{zhang2024holmesvad} and AnomalyRuler~\cite{yang2024anomalyruler} use multimodal instruction tuning or rule-based reasoning to localize and explain anomalies, while LAVAD~\cite{zanella2024lavad} and PLOVAD~\cite{xu2025plovad} further leverage captioning, prompting, and video-language interaction for anomaly detection. However, unlike real-world VAD which typically detects unusual human behaviors, anomalies in gameplay videos stem from violations of game mechanics, physics simulations, or rendering pipelines~\cite{hu2024gameagentsurvey, xu2024gameagentsurvey}. Consequently, video game glitch detection fundamentally differs from conventional VAD, requiring explicit reasoning over game-specific rules and virtual-world dynamics.

\section{\dataname}

% This section elaborates on the curation process of \textbf{\dataname}, a benchmark specifically made for open-ended video game glitch detection and temporal localization. The creation process is illustrated in Figure~\ref{fig:annotation}. It consists of three main stages: categorization, data filtering and balancing, and semi-automated annotation.

Figure~\ref{fig:annotation} illustrates the annotation pipeline of \textbf{\dataname}, which comprises 3 main stages: source video selection, pseudo glitch description generation, and human validation \& temporal grounding. We also provide detailed statistics in Table~\ref{tab:videoglitch_stats}.
%\lifu{categorization, data filtering and balancing are not really quite meaningful steps in this construction process, so please update the figure by (1) merge these two into one stage; (2) update the order of them based on the new description in 3.1; (3) we can still claim three stages, including source video selection, pseudo glitch description generation by GPT-4o, human validation and temporal grounding} \muyang{Done.}

\subsection{Selection of Source Videos}
Our dataset is built upon GamePhysics~\cite{clipxgamephysics}, a large-scale collection of 26,954 gameplay videos from the GamePhysics subreddit\footnote{\url{https://www.reddit.com/r/GamePhysics/}}
, where players frequently share community-reported glitches and unusual in-game events. GamePhysics provides the raw videos together with basic metadata, including game title (e.g., \texttt{Cyberpunk 2077}) and Reddit submission ID in filename. However, it was originally designed as a retrieval-oriented resource rather than a benchmark for open-ended glitch detection, so it does not provide verified glitch annotations, fine-grained natural-language glitch descriptions, or temporal boundaries indicating when a glitch occurs in the video.
% \lifu{confirm the description of GamePhysics here and add necessary additional information, so that reviewers can clearly understand how the original GamePhysics look like, what information has been included/provided by GamePhysics, and what information needs to be labeled by us in order to build our dataset} \muyang{Done.} 
We therefore use it as the source video pool and further annotate open-ended glitch descriptions and timestamps to construct \dataname. Although GamePhysics covers 1,873 unique games, many are represented by only one or two candidate videos without clear or confirmed glitches. To ensure sufficient coverage for reliable annotation and evaluation, we retain only games with at least ten candidate videos, yielding a final pool of 120 games.

Among these 120 games, the number of candidate videos varies, with some games contributing far more videos than others. To promote a balanced distribution of game types in \dataname{}, we first construct a taxonomy of genres and subgenres based on standard definitions from Wikipedia\footnote{\url{https://en.wikipedia.org/wiki/Action\_game}} and then use GPT-4o-mini~\cite{hurst2024gpt4o} to assign each game title to the corresponding predefined categories. The detailed taxonomy is provided in Appendix~\ref{appendix:stats}. Based on this taxonomy, we sample videos in a genre-aware manner, prioritizing underrepresented subgenres and games with fewer available videos. This sampling strategy improves diversity across genres, subgenres, and individual games, while reducing over-representation from a small number of dominant categories. As a result, we sample 5,238 gameplay videos spanning 120 games.

\begin{table}[!t]
\centering
\small
\caption{Key statistics of \dataname{}.
%\haibo{what about the average number of bugs in a single video?}\muyang{Done.}
}
\begin{tabular}{lc}
\toprule
\textbf{Statistics of \dataname} & \textbf{Value} \\
\midrule
Total number of videos &  5238\\
Game genres/subgenres &  6 / 21\\
\#Bugs in a single video (avg/max) &  1.03 / 6\\
Video length (seconds, avg/max) &  19.11 / 60.00\\
Bug description length (tokens, avg/max) &  35.94 / 164\\
\bottomrule
\end{tabular}
\label{tab:videoglitch_stats}
\end{table}

\subsection{Semi-Automated Annotation Pipeline}

Since purely manual annotation of fine-grained, temporally grounded glitches is prohibitively expensive at scale, we introduce a semi-automated pipeline. This approach synergizes MLLM-based video understanding with rigorous human validation to ensure both efficiency and high data quality.

\par \textbf{Automated generation of pseudo descriptions.} To reduce the annotation burden, we employ GPT-4o~\cite{hurst2024gpt4o} to generate pseudo glitch descriptions for each selected video. Because processing full-length gameplay videos can degrade the model's understanding, we partition each video into short segments ($<10$ seconds) and sample frames at 2 FPS, restricting the input to fewer than 20 frames per segment. To enhance generation quality, we augment the visual input with the original Reddit discussion context (i.e., post title and comments), retrieved via PRAW~\cite{praw_docs} using the submission ID embedded in each source filename. This supplementary metadata serves as a strong semantic prior, enabling the model to produce highly informative and accurate pseudo descriptions. An example of the model input and output is provided in Appendix~\ref{appendix:anno}.

\begin{figure*}[!t]
\centering
\includegraphics[width=\textwidth]{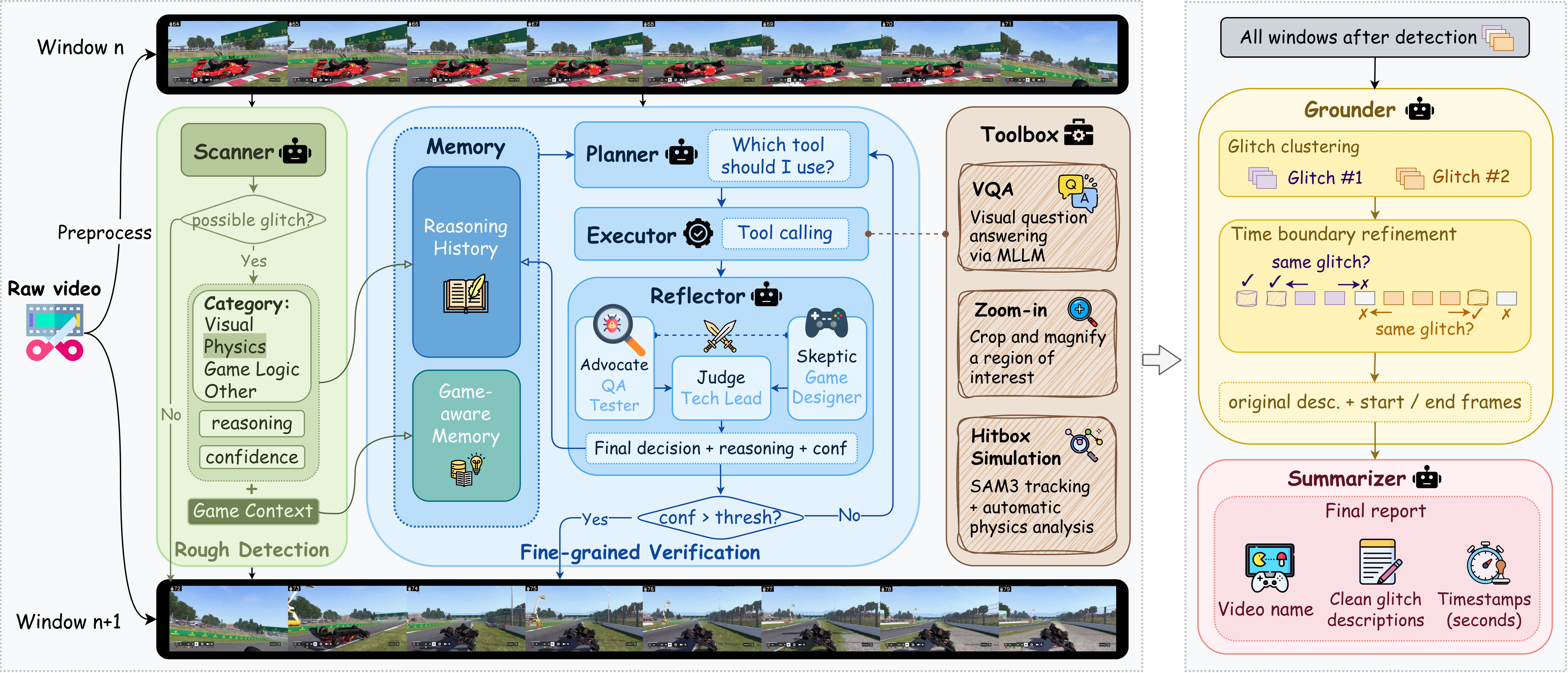}
\caption{Illustration of our \methodname{} framework.} \label{fig:framework}
\Description{Our pipeline.}
\end{figure*}

\par \textbf{Human validation and temporal grounding.}
% To guarantee the highest standard of data quality, the MLLM-generated pseudo descriptions underwent a strict human review process.
% We develop a multi-user annotation interface allowing annotators to process videos simultaneously.
% Annotators are tasked to verify the MLLM outputs, making necessary corrections or additions if the model missed critical elements or misidentified game characters. Also, they manually mark the \textbf{exact start and end timestamps} for each glitch event directly within the interface, providing the precise temporal boundaries required for advanced localization tasks. 
As MLLM-generated descriptions may omit critical details, misidentify game entities, or provide inaccurate event boundaries, all pseudo descriptions undergo rigorous human verification. To support efficient and scalable annotation, we develop a multi-user annotation interface that allows multiple annotators to inspect videos simultaneously. Annotators verify each MLLM-generated description and revise it when necessary, such as correcting factual errors, adding missing details, or refining references to game characters and objects. They also manually annotate the \textbf{exact start and end timestamps} of each glitch event within the interface, producing the precise temporal boundaries needed for temporal localization and grounding tasks. Since the pseudo descriptions are generated from short video segments, the intermediate annotations are initially segment-level. During human validation, annotators review the full video together with all segment-level pseudo descriptions, and manually merge those referring to the same underlying glitch into a unified video-level glitch annotation. If a glitch spans multiple adjacent segments or appears in multiple disjoint segments, annotators consolidate these observations into one glitch report and assign the corresponding temporal intervals at the video level.

% \lifu{for the second challenge we mentioned earlier, like one glitch may appear multiple times in a video discontinuously, will human annotators do anything? It's better to describe how we get such annotations here} \lifu{I mean because the annotation was for each segment, how we merge these segment-based annotations to the video-level, did the annotators manually merge them or are there any additional steps?} \muyang{Clarified.}

%\zihao{In the appendix, you should claim that human annotators have the professional knowledge of game glitch detection. Besides, you need to clarify whether human annotators are paid or volunteers to do this work.}  \haibo{You can also add the screenshot of the multi-user annotation interface in your appendix and demonstrate how annotators use it}\muyang{OK. annotator info \& interface in Appendix.}

\section{Methodology}

\subsection{Problem Formulation}

Given an input gameplay video $V$, the goal of open-ended video game glitch detection is to identify all glitch events that occur in the video, generate a natural-language description for each event, and localize its temporal extent. 
We formulate this task as a structured set prediction problem, where each glitch report is represented as a pair
$(d, T)$, with $d$ denotes a free-form glitch description and $T$ denotes its temporal span. Since the same glitch may recur multiple times throughout a video with gaps in between, $T$ may consist of one or more disjoint intervals. For each video, the ground-truth annotations are defined as a set of such reports $Y=\{(d_j^{gt}, T_j^{gt})\}_{j=1}^{M}$, and the model is required to predict a set of the same form $\hat{Y}=\{(d_i^{p}, T_i^{p})\}_{i=1}^{N}$, where $M$ and $N$ are the numbers of ground-truth and predicted glitch reports, respectively.
%. We use $Y=\{(d_j^{gt}, T_j^{gt})\}_{j=1}^{M}$ to denote the ground-truth set and $\hat{Y}=\{(d_i^{p}, T_i^{p})\}_{i=1}^{N}$ to denote the predicted set, where $M$ and $N$ are the numbers of ground-truth and predicted glitches, respectively.

%\haibo{Para.1 and Para.2 are actually the same thing. The formulation of $\mathcal{Y}$ and $\hat{\mathcal{Y}}$ is redundant. You don't need to write it twice.  Consider merging them by only describing the Para.2. (but keep the problem setting)} \muyang{Done.}

\subsection{The \methodname{} Framework}
\label{sec:overview}
We propose \methodname{}, an agentic framework for open-ended video game glitch detection with temporal localization. The task is challenging for three main reasons. \textbf{First}, determining whether an unusual event is truly a glitch often requires accumulated gameplay context over time, rather than inspection of a single suspicious frame. 
%\haibo{So the gameplay context here is a dynamic process instead of a static frame, which should be explicitly pointed out to emphasize what a gameplay context is}
\textbf{Second}, gameplay videos frequently contain rare yet valid in-game behaviors, such as abrupt camera shifts, exaggerated character poses during scripted animations, or sudden object movements caused by normal game mechanics; as a result, one-pass judgments are prone to false positives. %such as abrupt camera or object movements triggered by normal gameplay mechanics
%\haibo{briefly give examples of what are rare but valid in-game behaviors}
%, making single-pass judgments prone to false positives. 
\textbf{Third}, a real glitch may span multiple temporal windows or reappear intermittently, so local window-level detections can be fragmented and must be consolidated into complete event-level glitch intervals. %\haibo{so window-level detections can result in ..., and must be ...}

To address these challenges, \methodname{} is built around three core designs: a lightweight \textbf{game-aware memory} for preserving global gameplay context, a \textbf{debate-based verification mechanism} for distinguishing true glitches from unusual but valid behaviors, and an \textbf{event-level grounding strategy} for merging fragmented evidence into complete glitch events. These designs are instantiated in a five-stage pipeline, as illustrated in Figure~\ref{fig:framework}. 

\subsubsection{Preprocessing}
\label{sec:preprocess}
%Given an input gameplay video $V$, we first sample frames at rate $\tau$ fps and group them into non-overlapping windows of size $k$. The frames within each window are spatially stitched into a single composite image. This produces a window sequence 
% Given an input gameplay video $V$, we first sample frames at a rate of $\tau$ FPS and partition them into non-overlapping windows, each containing $k$ frames. The frames within each window are then spatially stitched into a single composite image. This yields a sequence of stitched windows $\mathcal{W}=\{w_j\}_{j=1}^{L}$, where $w_j$ denotes the $j$-th window and $L$ is the total number of windows. This step is motivated by efficiency. It preserves short-range temporal cues while reducing the number of MLLM calls, making later multi-step reasoning more efficient. 

Given an input gameplay video $V$ sampled at $\tau$ FPS, we partition it into non-overlapping $k$-frame windows. The frames within each window are spatially stitched into a composite image, yielding a sequence of stitched windows $\mathcal{W}=\{w_j\}_{j=1}^{L}$, where $w_j$ denotes the $j$-th window and $L$ is the total number of windows. This step efficiently preserves short-range temporal cues while minimizing MLLM calls for subsequent multi-step reasoning.

%\haibo{Is the index $_j$ means the $j$-th window? if so should be clarified. And what does the bracket \{ \} here mean.} \muyang{Clarified.}

\subsubsection{Initial Glitch Detection with Game-aware Memory}
\label{sec:scanner}
\methodname{} starts with a \textbf{Scanner} that processes each stitched window $w_j \in \mathcal{W}$ in a single pass. For each window, the Scanner predicts:
\[
(\hat{y}_j, c_j, s_j, q_j)=F_{\mathrm{scan}}(w_j),
\]
where $\hat{y}_j\in\{0,1\}$ indicates whether the window contains a potential glitch, $c_j$ denotes a coarse glitch category (e.g., visual, physics, or game logic), $s_j$ is a natural-language summary of the current gameplay context (e.g., scene type, visible entities, and ongoing behaviors), and $q_j\in[0,1]$ is the confidence score.

The binary predictions $\hat{y}_j$ are first used to filter out clearly normal windows and retain only a small set of candidate windows for deeper analysis. More importantly, the context descriptions $\{s_j\}_{j=1}^{L}$ are aggregated into a compact \textbf{game-aware memory}. Specifically, we apply an LLM-based summarization function over the window-level context descriptions to obtain a global summary context:
\[
M = G(\{s_j\}_{j=1}^{L}),
\]
where $M$ captures the overall gameplay scene, active entities, and ongoing dynamics in the video. This memory serves as a video-level contextual prior for downstream reasoning. Instead of judging each suspicious window in isolation, \methodname{} can compare local anomalies against the broader gameplay context, which helps determine whether an unusual event is actually inconsistent with the game state or simply part of normal gameplay.
%It helps the system judge whether a visually unusual event is consistent with the broader gameplay context or more likely to be a true glitch.

\begin{comment}
\haibo{This paragraph is a bit too colloquial and descriptive. We can formalize the inputs, outputs, and the memory update process using mathematical notation. For example, Let $w_j$ denote the $j$-th window of the video sequence. At each step $j$, we formulate this dual-task single-pass process as: $$(\hat{y}_j, c_j, s_j) = \mathcal{F}_{scan}(w_j, \mathcal{M}_{j-1})$$ where $\hat{y}_j \in \{0, 1\}$ is a binary indicator of whether a potential glitch is present, $c_j \in \mathcal{C}$ denotes the predicted coarse glitch category (e.g., visual, physics, or game logic), and $s_j$ is a natural language context summary describing the ongoing gameplay (e.g., scene type, visible entities, and behaviors). And the lightweight game-aware memory can be denoted as $\mathcal{M}_j$. The memory state is recursively updated via an integration function $\mathcal{U}$:
$$\mathcal{M}_j = \mathcal{U}(\mathcal{M}_{j-1}, s_j)$$
As the video is scanned, $\mathcal{M}_j$ dynamically accumulates contextual cues. And the notations in this section should be consistent with the notations in previous sections.} \muyang{Modified as suggested :D}
\end{comment}

\subsubsection{Fine-grained Verification via Debate-based Reasoning}
\label{sec:analyzer}
After initial glitch detection, each candidate window $w_j$ is further examined through a verification loop involving: a \textbf{Planner}, an \textbf{Executor}, and a \textbf{Reflector}. This stage will gather targeted evidence and determine whether the candidate truly corresponds to a glitch.
%For each candidate window $w_j$, \methodname{} performs a deeper investigation through an iterative agent loop with three components: a \textbf{Planner}, an \textbf{Executor}, and a \textbf{Reflector}. The goal of this stage is to collect more reliable evidence and verify whether the candidate window truly contains a glitch.

At verification step $t$, the Planner takes as input the current candidate window $w_j$, the Scanner's initial hypothesis, the game-aware memory $M$, and the cumulative investigation memory $I_{j,t-1}$. It then selects the next action according to a planning policy $\pi$:
\[
a_{j,t}=\pi(w_j, M, I_{j,t-1}),
\]
where $a_{j,t}$ specifies both the tool to invoke and its associated arguments. %Our toolbox includes three operations
The available tools include: \textbf{VQA}, which asks targeted visual questions about the stitched window based on the current glitch hypothesis and the missing evidence needed for verification (e.g., ``\textit{How does the robotic arm interact with the wall and floor structures between frames \#24 and \#31?}''); \textbf{Zoom-in}, which selects a local image region for magnified inspection; and \textbf{Object Tracking}, which provides a short target description and invokes a segmentation-based tracker (SAM3~\cite{carion2025sam}) to obtain motion evidence. The Executor applies $a_{j,t}$ and returns a new observation $o_{j,t}$, which is then appended to the investigation memory.

After each tool execution, the resulting observation is evaluated by the Reflector through a structured debate among three roles: an \textbf{Advocate} (game QA tester), a \textbf{Skeptic} (game designer), and a \textbf{Judge} (tech lead). The Advocate argues that the observed phenomenon is a genuine glitch, while the Skeptic proposes plausible in-game explanations or highlights missing evidence. The Judge then arbitrates between the two sides and produces a binary verdict: $v_{j,t}\in\{\texttt{glitch},\texttt{normal}\}$, together with a confidence score $conf_{j,t}\in[0,1]$. This debate-based verification is a core design of \methodname{}. Rather than merely re-checking a prediction, it explicitly contrasts a glitch hypothesis against alternative explanations grounded in normal gameplay behavior. This matters because unusual motions or visual effects may still be valid under the game rules, such as exaggerated character poses during specific animations or abrupt object movements triggered by normal game mechanics. By forcing the system to reason over competing interpretations before making a decision, \methodname{} reduces false positives and improves verification reliability. %By explicitly considering these competing interpretations before making a decision, \methodname{} improves verification reliability and reduces false positives caused by visually unusual but valid gameplay.

The verification loop terminates when the Judge outputs a verdict with confidence above a threshold $\tau$, or when the maximum number of steps $T_{\max}$ is reached: 
$conf_{j,t}\ge\tau$ or $t=T_{\max}.$
The final verified result for window $w_j$ is then forwarded to the next stage.

\begin{comment}
\haibo{Still, we can formalize the iterative verification process using mathematical notation. For example, let $t$ denote the current step of the verification loop for window $w_j$. At step $t$, the Planner determines the next action $a_t$ based on a policy $\pi$:
$$a_t = \pi(\mathcal{I}_{t-1}, \mathcal{M}_j)$$
where $\mathcal{I}_{t-1}$ is the cumulative investigation memory up to step $t-1$, and $\mathcal{M}_j$ is the game-aware memory. We should explicitly state that $a_t$ includes both the tool selection and normalized 2D coordinates to guide tools like SAM3. The Executor then applies $a_t$ to yield a new observation $o_t$. The Reflector evaluates $o_t$ through a multi-role debate, where the Judge arbitrates to output a binary verdict $v_t \in \{0, 1\}$ and a confidence score $conf_t \in [0, 1]$. The iterative loop terminates early if $conf_t \ge \tau$ (where $\tau$ is a predefined threshold) or when the maximum step limit $T_{max}$ is reached. And these step indices and notations should also consistent with the rest of the paper. Since this is a core part of your method, you can also post a pseud-code here}

\haibo{Maybe I have something wrong in the algorithm details, but the core point is the same that we can use some mathematical notations to desribe the method instead of full language description} \muyang{Thanks for the examples! Tried my best to rewrite it.}
\end{comment}

\subsubsection{Event-level Grounding}
\label{sec:grounder}
The verification stage operates at the window level, whereas the target output requires event-level glitch reports with complete temporal spans. To bridge this gap, \methodname{} includes an \textbf{event-level Grounder} that consolidates fragmented window evidence into coherent glitch events.

The Grounder proceeds in two steps. First, it performs \textbf{semantic clustering} over the verified glitch windows. For each pair of verified windows, an LLM judges whether their descriptions refer to the same underlying glitch phenomenon. Windows deemed semantically consistent are grouped into the same event cluster. This allows the framework to handle cases where the same glitch appears across non-consecutive portions of the video.
Second, for each event cluster, \methodname{} performs \textbf{bidirectional temporal propagation} to refine the event boundaries. Starting from the initially detected windows, the model iteratively checks neighboring windows in both temporal directions and extends the boundaries whenever the same glitch remains visible. In this way, the Grounder can recover the full temporal extent of a glitch even when the initial detector only captures its most salient moments.
%The verification stage produces decisions at the window level, but the target output of our task is an event-level glitch report with complete temporal intervals. To bridge this gap, \methodname{} includes an \textbf{event-level Grounder} that merges fragmented window evidence into coherent glitch events.
%\par The Grounder operates in two steps. First, it performs \textbf{semantic clustering} over verified glitch windows. For each pair of verified windows, we use an LLM to judge whether their descriptions refer to the same underlying glitch. Windows judged as semantically consistent are grouped into the same event cluster, allowing the system to handle cases where one glitch appears across non-consecutive segments.
%\haibo{How does the semantic clustering work? It's a Sentence-BERT? Or you just use the LLM? Should be clarified} \muyang{LLM. Clarified.} 
%Second, for each cluster, \methodname{} performs \textbf{bidirectional temporal propagation} to refine its start and end boundaries. Starting from the detected windows, the model checks neighboring windows in both directions and extends the boundary when the same glitch is still visible. This design could recover full glitch intervals even when the initial detector only captures the most salient part of the event.

\subsubsection{Structured Report Generation}
\label{sec:summarizer}
Finally, \methodname{} converts the grounded event clusters into the output set $\hat{Y}$. Since a single gameplay video may contain multiple distinct glitches, the framework generates a set of structured reports rather than a single prediction. For each cluster, it summarizes the accumulated multi-window evidence into a single coherent description and converts the refined frame ranges into timestamp intervals in seconds. The final output is therefore a structured glitch report containing natural-language description together with one or more temporal spans.

% \lifu{at somewhere, we need to emphasize that there could be multiple glitches from a single video} \muyang{Done.}

%Finally, \methodname{} converts the grounded event clusters into the output set $\hat{Y}$. For each event, the model summarizes the accumulated window-level evidence into a single coherent natural-language glitch description and converts the refined frame ranges into timestamp intervals in seconds. The final output is a structured glitch report containing a free-form description together with one or more temporal spans, which can be directly evaluated under the protocol in Section~\ref{sec:eval-criteria}.

\begin{table*}[ht]
\centering
\small
\setlength{\tabcolsep}{6pt}
\caption{Comparison of vanilla models and \methodname{}-enhanced models on \dataname{} across description generation, temporal grounding, and overall performance.}
\label{tab:main_results}
\begin{tabular}{lccccc}
\toprule
\multirow{2}{*}{\textbf{Model}} & \multicolumn{3}{c}{\textbf{Description Generation}} & \textbf{Temporal Grounding} & \textbf{Overall} \\
\cmidrule(lr){2-4} \cmidrule(lr){5-5} \cmidrule(lr){6-6}
& \textbf{Precision (\%)} & \textbf{Recall (\%)} & \textbf{F1 (\%)} & \textbf{mIoU} & \textbf{F1 $\times$ IoU (\%)} \\
\midrule

\rowcolor{gray!15}
\multicolumn{6}{l}{\textbf{Proprietary Models}} \\
\midrule
gemini-2.0-flash~\cite{gemini2} & 18.36 & 25.35 & 21.29 & 0.44 & 10.55 \\
gpt-4o-mini~\cite{hurst2024gpt4o} & 12.33 & 32.36 & 17.86 & 0.35 & 6.20 \\
claude-3.5-haiku~\cite{claude3.5} & 21.66 & 32.54 & 26.01 & 0.47 & 12.91 \\
nova-lite-v1~\cite{nova} & 6.66 & 23.77 & 10.41 & 0.28 & 2.98 \\

\midrule
\rowcolor{gray!15}
\multicolumn{6}{l}{\textbf{Open-source Models}} \\
\midrule
Qwen2.5-VL-3B-Instruct~\cite{Qwen2.5-VL} & 11.08 & 26.53 & 15.63 & 0.31 & 4.81 \\
\textbf{+\methodname{}} & \textbf{32.36} (+21.28) & \textbf{39.38} (+12.85) & \textbf{35.52} (+19.89) & \textbf{0.48} (+0.17) & \textbf{17.29} (+12.48) \\

Qwen2.5-VL-7B-Instruct~\cite{Qwen2.5-VL} & 10.41 & 14.74 & 12.20 & 0.30 & 4.11 \\
\textbf{+\methodname{}} & \textbf{34.55} (+24.14) & \textbf{45.09} (+30.35) & \textbf{39.12} (+26.92) & \textbf{0.53} (+0.23) & \textbf{19.46} (+15.35) \\

InternVL2.5-4B~\cite{intern2.5vl} & 8.62 & 22.93 & 12.53 & 0.18 & 1.92 \\
\textbf{+\methodname{}} & \textbf{29.15} (+20.53) & \textbf{36.87} (+13.94) & \textbf{32.56} (+20.03) & \textbf{0.52} (+0.34) & \textbf{15.27} (+13.35) \\

InternVL2.5-8B~\cite{intern2.5vl} & 11.90 & 23.36 & 15.77 & 0.25 & 4.06 \\
\textbf{+\methodname{}} & \textbf{32.64} (+20.74) & \textbf{40.97} (+17.61) & \textbf{36.33} (+20.56) & \textbf{0.50} (+0.25) & \textbf{17.04} (+12.98) \\

UI-TARS-1.5-7B~\cite{ui-tars-15-seed} & 19.08 & 24.93 & 21.62 & 0.40 & 9.11 \\
\textbf{+\methodname{}} & \textbf{34.31} (+15.23) & \textbf{49.28} (+24.35) & \textbf{40.45} (+18.83) & \textbf{0.51} (+0.11) & \textbf{17.02} (+7.91) \\

LLaVA-OneVision-7B~\cite{llavaonevision} & 5.90 & 19.50 & 9.06 & 0.26 & 2.18 \\
\textbf{+\methodname{}} & \textbf{30.13} (+24.23) & \textbf{34.85} (+15.35) & \textbf{32.32} (+23.26) & \textbf{0.50} (+0.24) & \textbf{16.24} (+14.06) \\

\bottomrule
\end{tabular}
\end{table*}

\section{Experiments}

\subsection{Experimental Setup}
We evaluate a diverse set of multimodal models, including both proprietary and open-source backbones. The proprietary models are gemini-2.0-flash~\cite{gemini2}, gpt-4o-mini~\cite{hurst2024gpt4o}, claude-3.5-haiku~\cite{claude3.5}, and nova-lite-v1~\cite{nova}. The open-source models are Qwen2.5-VL-3B/7B-Instruct~\cite{Qwen2.5-VL}, InternVL2.5-4B/8B~\cite{intern2.5vl}, UI-TARS-1.5-7B~\cite{ui-tars-15-seed}, and LLaVA-OneVision-7B~\cite{llavaonevision}. Our \methodname{} framework comparison is conducted on open-source models, while proprietary models are included as reference baselines. Videos are uniformly sampled at 4 FPS and divided into non-overlapping windows of 8 frames. The frames within each window are spatially stitched into a single composite image before being fed into the model. All experiments are run on four NVIDIA Quadro RTX 8000 GPUs (48GB). Please refer to Appendix~\ref{appendix:detailed_setup} for detailed experiment and evaluation setup.

\subsection{Evaluation Protocol}
\label{sec:eval-criteria}

%\haibo{For me, the evaluation criteria section is too long, and its length is even longer than that of your method section and benchmark section. You only need to briefly describe your evaluation criteria here, because this is not your major contribution, and the details can be in the appendix.} \muyang{OK I'm trying to compress}
% \lifu{update this paragraph to include all the evaluation metrics for description generation, temporal grounding, and overall. In the following stages, some are about F-score of description generation, so first define it, then the overall is a combination of F-score and mIoU} \muyang{Done.}
Open-ended video game glitch detection requires evaluating both semantic correctness and temporal localization. Since prior game glitch benchmarks mainly focus on multiple-choice accuracy or description-only quality, we design a task-specific evaluation protocol by building on SODA~\cite{dvcsurvey} and adapting it to our setting. For one video, let the predicted set be $\hat{Y}=\{(d_i^p,T_i^p)\}_{i=1}^N$ and the ground-truth set be $Y=\{(d_j^{gt},T_j^{gt})\}_{j=1}^M$.
\par \textbf{Stage 1: LLM-as-Judge Semantic Scoring.}
For each prediction--ground-truth pair, we use gpt-4o-mini to score the semantic similarity between the predicted description $d_i^p$ and the ground-truth description $d_j^{gt}$, denoted as $S_{\mathrm{LLM}}(d_i^p,d_j^{gt}) \in [0,5]$, where a higher score indicates better semantic alignment.

\par \textbf{Stage 2: LLM-Weighted IoU Computation.} 
To support reliable matching, we combine semantic similarity and temporal overlap into a joint score, $W_{i,j}=S_{\mathrm{LLM}}(d_i^p,d_j^{gt}) \cdot \mathrm{IoU}(T_i^p,T_j^{gt})$. This score is high only when a prediction matches a ground-truth glitch in both description and temporal span.

\par \textbf{Stage 3: Matching.} 
Given the pairwise score matrix $W \in \mathbb{R}^{N \times M}$, we perform global one-to-one matching with the Hungarian algorithm~\cite{kuhn1955hungarian}. Let the matched pairs be denoted by $\mathcal{A}=\{(i,j)\}$.

\par \textbf{Stage 4: Final Metrics Computation.} 
Based on $\mathcal{A}$, we report three groups of metrics as follows.

\emph{Description generation.}
We first measure semantic quality independently. Precision is $P_{\mathrm{desc}}=\sum_{(i,j)\in\mathcal{A}} S_{\mathrm{LLM}}(d_i^p,d_j^{gt})/N$, and recall is $R_{\mathrm{desc}}=\sum_{(i,j)\in\mathcal{A}} S_{\mathrm{LLM}}(d_i^p,d_j^{gt})/M$. Their harmonic mean~\cite{schutze2008f1} is reported as $F1_{\mathrm{desc}}$.

\emph{Temporal Grounding.}
We use the mean IoU over matched pairs, defined as $\mathrm{mIoU}=\sum_{(i,j)\in\mathcal{A}} \mathrm{IoU}(T_i^p,T_j^{gt})/|\mathcal{A}|$.

\emph{Overall Performance.}
Finally, we evaluate semantic correctness and temporal localization jointly. We compute Precision as $P_{\mathrm{overall}}=\sum_{(i,j)\in\mathcal{A}} W_{i,j}/N$ and recall as $R_{\mathrm{overall}}=\sum_{(i,j)\in\mathcal{A}} W_{i,j}/M$, and report their harmonic mean as $F1 \times \mathrm{IoU}$.

\begin{table}[!t]
\centering
\small
\caption{Ablation study on the game-aware memory and debate-based verification mechanism.}
\label{tab:ablation_desc}
\begin{tabular}{lccc}
\toprule
\textbf{Variant} & \textbf{Precision (\%)} & \textbf{Recall (\%)} & \textbf{F1 (\%)} \\
\midrule
\methodname{} & 34.55 & 45.09 & 39.12 \\
w/o Game-aware memory & 30.65 & 35.80 & 33.03 \\
w/o Debate-based verification & 28.94 & 36.21 & 32.17 \\
\bottomrule
\end{tabular}
\end{table}

\begin{table}[!t]
\centering
\small
\caption{Ablation study on the event-level grounding strategy across different open-source backbones (mIoU).}
\label{tab:ablation_grounding}
\begin{tabular}{lcc}
\toprule
\textbf{Backbone} & \textbf{\methodname{}} & \textbf{w/o Event-level grounding} \\
\midrule
Qwen2.5-VL-7B-Instruct & 0.53 & 0.32 \\
InternVL2.5-8B & 0.50 & 0.24 \\
UI-TARS-1.5-7B & 0.51 & 0.36 \\
\bottomrule
\end{tabular}
\end{table}

\subsection{Main Results}
Table~\ref{tab:main_results} reports the main evaluation results on \dataname. From these results, we summarize the following observations.
\par \textbf{Current models still struggle on open-ended video game glitch detection.}
Even strong multimodal models achieve only limited performance on this task. For proprietary baselines, the best F1 is only 26.01\% (claude-3.5-haiku), with an mIoU of 0.47. For open-source baselines, the best F1 is 21.62\% (UI-TARS-1.5-7B), and the average F1 across all six open-source models is only 14.47\%, with an average mIoU of 0.28. These results suggest that \dataname{} remains highly challenging for current MLLMs, especially when models must both describe glitches in open-ended language and localize them precisely in time.
\par \textbf{\methodname{} consistently improves both description generation and temporal grounding.}
After applying \methodname{}, all six open-source backbones show clear gains on all major metrics. Averaged over six open-source models, F1 improves from 14.47\% to 36.05\% (+21.58\%), while mIoU improves from 0.28 to 0.51 (+0.23). The overall F1$\times$IoU score also rises from 4.37\% to 17.05\% on average, indicating that the improvements are not limited to one aspect of the task. Notably, after applying \methodname{}, all open-source models achieve higher F1, mIoU and F1$\times$IoU scores than the proprietary baselines reported in Table~\ref{tab:main_results}. This shows that the proposed framework brings robust improvements across different model families and improves both semantic accuracy and temporal completeness.

\begin{figure*}[!t]
\centering
\includegraphics[width=0.95\textwidth]{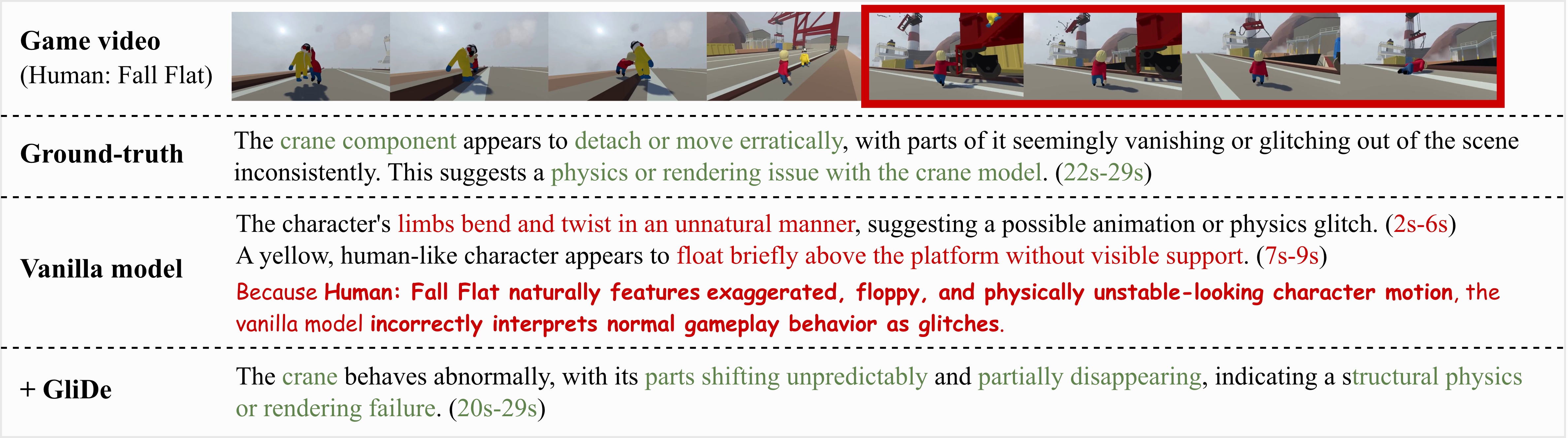}
\caption{Qualitative comparison between the vanilla model and \methodname{} on a \texttt{Human: Fall Flat} gameplay clip.}
\label{fig:case_human}
\Description{Case analysis.}
\end{figure*}

\begin{figure*}[!t]
\centering
\includegraphics[width=0.95\textwidth]{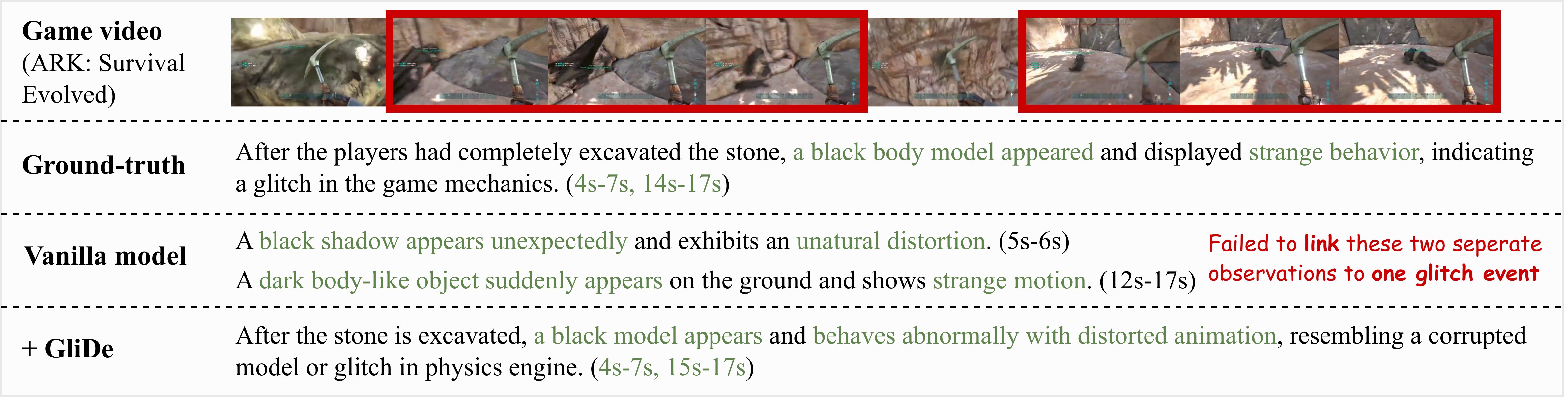}
\caption{Qualitative comparison between the vanilla model and \methodname{} on a \texttt{ARK: Survival Evolved} gameplay clip.}
\label{fig:case_ark}
\Description{Case analysis.}
\end{figure*}

\par \textbf{Case analysis.} We present two examples to illustrate the two core challenges of \dataname{}. The first case, shown in Figure~\ref{fig:case_human}, highlights the difficulty of distinguishing genuine glitches from visually unusual but valid in-game behaviors. In \texttt{Human: Fall Flat}, the vanilla model is distracted by the game's exaggerated and floppy character motions, falsely identifying twisted limbs and brief apparent floating as glitches, while missing the actual anomaly in the final segment. In contrast, \methodname{} focuses on the true crane-related failure, producing a prediction that is much closer to the ground truth. The second case in Figure~\ref{fig:case_ark} highlights the temporal challenge of fragmented and repeated glitches. In \texttt{ARK: Survival Evolved}, the same black model anomaly appears in two disjoint intervals. The vanilla model detects the two segments separately and fails to associate them as one glitch event. By contrast, \methodname{} links the repeated observations through semantically consistent evidence and produces a single coherent glitch report with multiple temporal spans. These cases show that \methodname{} improves both context-aware verification and event-level temporal grounding.

\begin{figure}[!t]
    \centering
    \includegraphics[width=0.95\columnwidth]{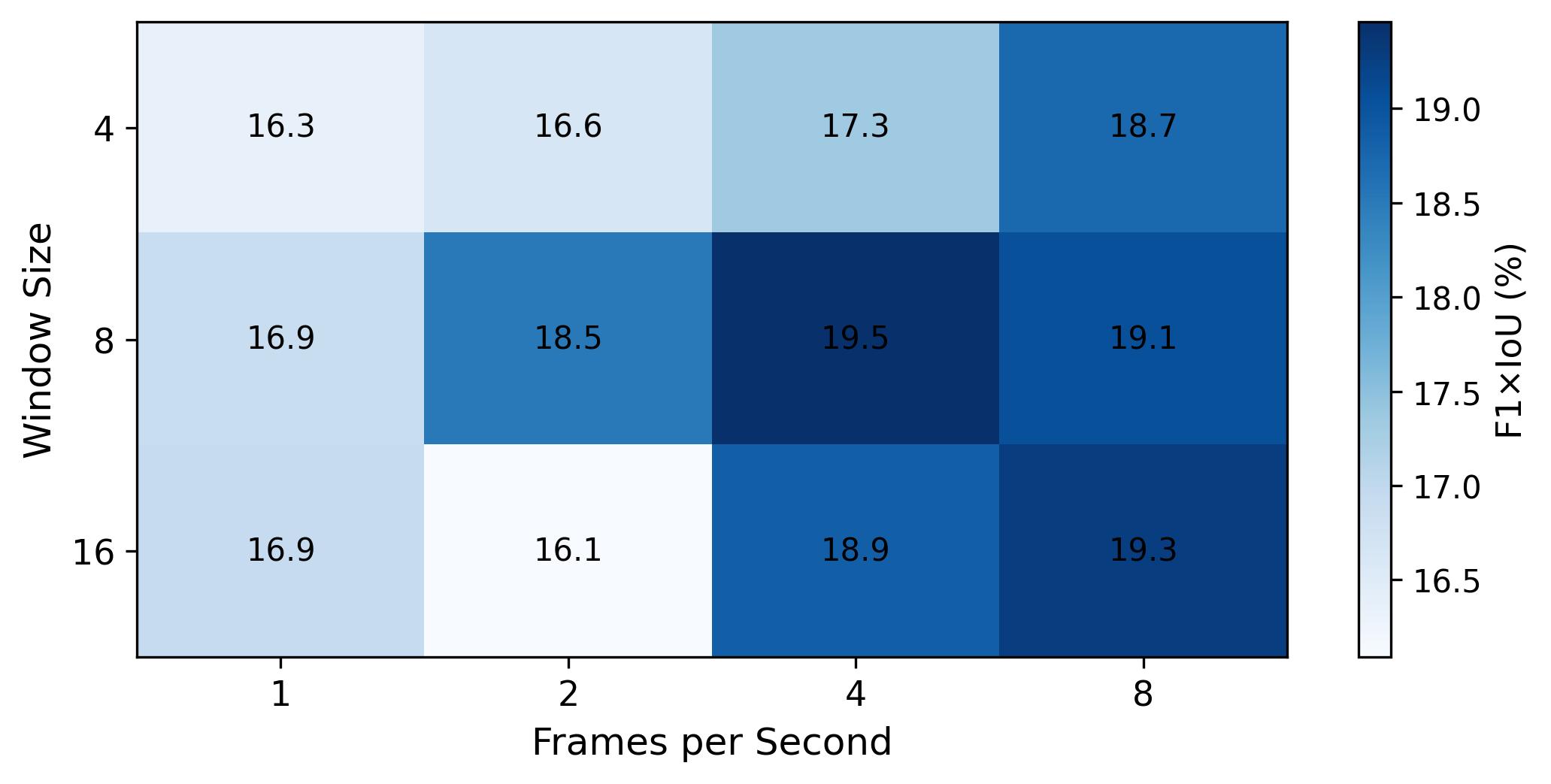}
    \caption{Hyperparameter sensitivity study.}
    \label{fig:hyperparameter}
\end{figure}

\subsection{Ablation Studies}
We conduct ablation studies on the three key designs in \methodname{}: game-aware memory, debate-based verification mechanism, and event-level grounding strategy, as shown in Tables~\ref{tab:ablation_desc} and ~\ref{tab:ablation_grounding}. 

Table~\ref{tab:ablation_desc} reports the ablation results on Qwen2.5-VL-7B-Instruct. In particular, removing Game context reduces F1 from 39.12\% to 33.03\%, showing that game-aware memory help the model better understand ongoing gameplay and recognize valid glitches. Removing debate-based verification mechanism causes a further drop in precision, from 34.55\% to 28.94\%, indicating that the debate mechanism is especially useful for filtering unreliable glitch predictions and reducing false positives.

Table~\ref{tab:ablation_grounding} shows that the event-level grounding strategy consistently improves temporal localization accuracy across three open-source models. After removing this strategy, mIoU drops significantly across all backbones, indicating that event-level grounding is essential for recovering accurate glitch intervals. This is particularly important in our setting, where a single glitch may span multiple windows or reappear across disjoint time intervals.

Furthermore, we conduct a hyperparameter sensitivity study on FPS and window size (on Qwen2.5-VL-7B-Instruct). Figure~\ref{fig:hyperparameter} shows that \methodname{} remains relatively stable across different settings, while moderate temporal granularity generally gives slightly better results. The best performance is obtained at 4 FPS with a window size of 8, suggesting that this combination provides a good balance between temporal detail and contextual coverage. When the FPS is too low, short glitch evidence may be missed, while overly large window sizes can dilute local glitch signals and make reasoning less focused. These results support the default preprocessing choice used in our experiments.

\section{Conclusion}

In this paper, we introduce \dataname, a benchmark for open-ended video game glitch detection with temporal grounding. We also propose \methodname{}, an agentic framework designed for this task. \dataname{} requires models to detect glitches from raw gameplay videos, describe them in natural language, and localize when they occur. To address this challenge, \methodname{} combines game-aware context, debate-based verification, and event-level grounding. Experiments show that current multimodal models still struggle on this task, while \methodname{} improves both description quality and temporal localization. We hope this work can support future research on game video understanding and agentic multimodal reasoning.

\bibliographystyle{ACM-Reference-Format}
\bibliography{references}

@inproceedings{clipxgamephysics,
  title={Clip meets gamephysics: Towards bug identification in gameplay videos using zero-shot transfer learning},
  author={Taesiri, Mohammad Reza and Macklon, Finlay and Bezemer, Cor-Paul},
  booktitle={Proceedings of the 19th International Conference on Mining Software Repositories},
  pages={270--281},
  year={2022}
}

@article{gamebugdesc,
  title={Large language models are pretty good zero-shot video game bug detectors},
  author={Taesiri, Mohammad Reza and Macklon, Finlay and Wang, Yihe and Shen, Hengshuo and Bezemer, Cor-Paul},
  journal={arXiv preprint arXiv:2210.02506},
  year={2022}
}

@inproceedings{glitchbench,
  title={GlitchBench: Can large multimodal models detect video game glitches?},
  author={Taesiri, Mohammad Reza and Feng, Tianjun and Bezemer, Cor-Paul and Nguyen, Anh},
  booktitle={Proceedings of the IEEE/CVF Conference on Computer Vision and Pattern Recognition},
  pages={22444--22455},
  year={2024}
}

@inproceedings{videogamebunny,
  title={Videogamebunny: Towards vision assistants for video games},
  author={Taesiri, Mohammad Reza and Bezemer, Cor-Paul},
  booktitle={2025 IEEE/CVF Winter Conference on Applications of Computer Vision (WACV)},
  pages={1403--1413},
  year={2025},
  organization={IEEE}
}

@article{physvlm,
  title={Physgame: Uncovering physical commonsense violations in gameplay videos},
  author={Cao, Meng and Tang, Haoran and Zhao, Haoze and Guo, Hangyu and Liu, Jiaheng and Zhang, Ge and Liu, Ruyang and Sun, Qiang and Reid, Ian and Liang, Xiaodan},
  journal={arXiv preprint arXiv:2412.01800},
  year={2024}
}

@article{physgame,
  title={Order from Chaos: Physical World Understanding from Glitchy Gameplay Videos},
  author={Cao, Meng and Tang, Haoran and Zhao, Haoze and Han, Mingfei and Liu, Ruyang and Sun, Qiang and Chang, Xiaojun and Reid, Ian and Liang, Xiaodan},
  journal={arXiv preprint arXiv:2601.16471},
  year={2026}
}

@inproceedings{taesirivideogameqa,
  title={VideoGameQA-Bench: Evaluating Vision-Language Models for Video Game Quality Assurance},
  author={Taesiri, Mohammad Reza and Ghildyal, Abhijay and Zadtootaghaj, Saman and Barman, Nabajeet and Bezemer, Cor-Paul},
  booktitle={The Thirty-ninth Annual Conference on Neural Information Processing Systems Datasets and Benchmarks Track}
}

@inproceedings{sultani2018ucfcrime,
  title={Real-world anomaly detection in surveillance videos},
  author={Sultani, Waqas and Chen, Chen and Shah, Mubarak},
  booktitle={Proceedings of the IEEE conference on computer vision and pattern recognition},
  pages={6479--6488},
  year={2018}
}

@inproceedings{luo2017shanghaitech,
  title={A revisit of sparse coding based anomaly detection in stacked rnn framework},
  author={Luo, Weixin and Liu, Wen and Gao, Shenghua},
  booktitle={Proceedings of the IEEE international conference on computer vision},
  pages={341--349},
  year={2017}
}

@inproceedings{wu2020xdviolence,
  title={Not only look, but also listen: Learning multimodal violence detection under weak supervision},
  author={Wu, Peng and Liu, Jing and Shi, Yujia and Sun, Yujia and Shao, Fangtao and Wu, Zhaoyang and Yang, Zhiwei},
  booktitle={European conference on computer vision},
  pages={322--339},
  year={2020},
  organization={Springer}
}

@article{tang2025videosurvey,
  title={Video understanding with large language models: A survey},
  author={Tang, Yunlong and Bi, Jing and Xu, Siting and Song, Luchuan and Liang, Susan and Wang, Teng and Zhang, Daoan and An, Jie and Lin, Jingyang and Zhu, Rongyi and others},
  journal={IEEE Transactions on Circuits and Systems for Video Technology},
  year={2025},
  publisher={IEEE}
}

@inproceedings{wang2024videoagent,
  title={Videoagent: Long-form video understanding with large language model as agent},
  author={Wang, Xiaohan and Zhang, Yuhui and Zohar, Orr and Yeung-Levy, Serena},
  booktitle={European Conference on Computer Vision},
  pages={58--76},
  year={2024},
  organization={Springer}
}

@inproceedings{shang2024traveler,
  title={TraveLER: A Modular Multi-LMM Agent Framework for Video Question-Answering},
  author={Shang, Chuyi and You, Amos and Subramanian, Sanjay and Darrell, Trevor and Herzig, Roei},
  booktitle={Proceedings of the 2024 Conference on Empirical Methods in Natural Language Processing},
  pages={9740--9766},
  year={2024}
}

@inproceedings{liu2025videomind,
  title={VideoMind: A Chain-of-LoRA Agent for Temporal-Grounded Video Reasoning},
  author={Liu, Ye and Lin, Kevin Qinghong and Chen, Chang Wen and Shou, Mike Zheng},
  booktitle={NeurIPS 2025 Workshop on Bridging Language, Agent, and World Models for Reasoning and Planning}
}

@article{sestini2022ccpt,
  title={Automated gameplay testing and validation with curiosity-conditioned proximal trajectories},
  author={Sestini, Alessandro and Gissl{\'e}n, Linus and Bergdahl, Joakim and Tollmar, Konrad and Bagdanov, Andrew David},
  journal={IEEE Transactions on Games},
  volume={16},
  number={1},
  pages={113--126},
  year={2022},
  publisher={IEEE}
}

@inproceedings{liu2022inspector,
  title={Inspector: Pixel-based automated game testing via exploration, detection, and investigation},
  author={Liu, Guoqing and Cai, Mengzhang and Zhao, Li and Qin, Tao and Brown, Adrian and Bischoff, Jimmy and Liu, Tie-Yan},
  booktitle={2022 IEEE Conference on Games (CoG)},
  pages={237--244},
  year={2022},
  organization={IEEE}
}

@article{autoplaysurvey,
  title={Assessing adaptive world models in machines with novel games},
  author={Ying, Lance and Collins, Katherine M and Sharma, Prafull and Colas, Cedric and Zhao, Kaiya Ivy and Weller, Adrian and Tavares, Zenna and Isola, Phillip and Gershman, Samuel J and Andreas, Jacob D and others},
  journal={arXiv preprint arXiv:2507.12821},
  year={2025}
}

@article{ariyurek2019gametestagents,
  title={Automated video game testing using synthetic and humanlike agents},
  author={Ariyurek, Sinan and Betin-Can, Aysu and Surer, Elif},
  journal={IEEE Transactions on Games},
  volume={13},
  number={1},
  pages={50--67},
  year={2019},
  publisher={IEEE}
}

@inproceedings{bergdahl2020drlgametest,
  title={Augmenting automated game testing with deep reinforcement learning},
  author={Bergdahl, Joakim and Gordillo, Camilo and Tollmar, Konrad and Gissl{\'e}n, Linus},
  booktitle={2020 IEEE Conference on Games (CoG)},
  pages={600--603},
  year={2020},
  organization={IEEE}
}

@inproceedings{sultani2018vadsurvey,
  title={Real-world anomaly detection in surveillance videos},
  author={Sultani, Waqas and Chen, Chen and Shah, Mubarak},
  booktitle={Proceedings of the IEEE conference on computer vision and pattern recognition},
  pages={6479--6488},
  year={2018}
}

@article{ramachandra2020vadsurvey,
  title={A survey of single-scene video anomaly detection},
  author={Ramachandra, Bharathkumar and Jones, Michael J and Vatsavai, Ranga Raju},
  journal={IEEE transactions on pattern analysis and machine intelligence},
  volume={44},
  number={5},
  pages={2293--2312},
  year={2020},
  publisher={IEEE}
}

@inproceedings{hasan2016dnnvad,
  title={Learning temporal regularity in video sequences},
  author={Hasan, Mahmudul and Choi, Jonghyun and Neumann, Jan and Roy-Chowdhury, Amit K and Davis, Larry S},
  booktitle={Proceedings of the IEEE conference on computer vision and pattern recognition},
  pages={733--742},
  year={2016}
}

@inproceedings{liu2018dnnvad,
  title={Future frame prediction for anomaly detection--a new baseline},
  author={Liu, Wen and Luo, Weixin and Lian, Dongze and Gao, Shenghua},
  booktitle={Proceedings of the IEEE conference on computer vision and pattern recognition},
  pages={6536--6545},
  year={2018}
}

@inproceedings{gong2019dnnvad,
  title={Memorizing normality to detect anomaly: Memory-augmented deep autoencoder for unsupervised anomaly detection},
  author={Gong, Dong and Liu, Lingqiao and Le, Vuong and Saha, Budhaditya and Mansour, Moussa Reda and Venkatesh, Svetha and Hengel, Anton van den},
  booktitle={Proceedings of the IEEE/CVF international conference on computer vision},
  pages={1705--1714},
  year={2019}
}

@article{zhang2024holmesvad,
  title={Holmes-vad: Towards unbiased and explainable video anomaly detection via multi-modal llm},
  author={Zhang, Huaxin and Xu, Xiaohao and Wang, Xiang and Zuo, Jialong and Han, Chuchu and Huang, Xiaonan and Gao, Changxin and Wang, Yuehuan and Sang, Nong},
  journal={arXiv preprint arXiv:2406.12235},
  year={2024}
}

@inproceedings{yang2024anomalyruler,
  title={Follow the rules: reasoning for video anomaly detection with large language models},
  author={Yang, Yuchen and Lee, Kwonjoon and Dariush, Behzad and Cao, Yinzhi and Lo, Shao-Yuan},
  booktitle={European Conference on Computer Vision},
  pages={304--322},
  year={2024},
  organization={Springer}
}

@inproceedings{zanella2024lavad,
  title={Harnessing large language models for training-free video anomaly detection},
  author={Zanella, Luca and Menapace, Willi and Mancini, Massimiliano and Wang, Yiming and Ricci, Elisa},
  booktitle={Proceedings of the IEEE/CVF Conference on Computer Vision and Pattern Recognition},
  pages={18527--18536},
  year={2024}
}

@article{xu2025plovad,
  title={Plovad: Prompting vision-language models for open vocabulary video anomaly detection},
  author={Xu, Chenting and Xu, Ke and Jiang, Xinghao and Sun, Tanfeng},
  journal={IEEE Transactions on Circuits and Systems for Video Technology},
  volume={35},
  number={6},
  pages={5925--5938},
  year={2025},
  publisher={IEEE}
}

@article{lin2019identifybugs,
  title={Identifying gameplay videos that exhibit bugs in computer games},
  author={Lin, Dayi and Bezemer, Cor-Paul and Hassan, Ahmed E},
  journal={Empirical Software Engineering},
  volume={24},
  number={6},
  pages={4006--4033},
  year={2019},
  publisher={Springer}
}

@article{wilkins20223dbugs,
  title={Learning to identify perceptual bugs in 3d video games},
  author={Wilkins, Benedict and Stathis, Kostas},
  journal={arXiv preprint arXiv:2202.12884},
  year={2022}
}

@article{hu2024gameagentsurvey,
  title={A survey on large language model-based game agents},
  author={Hu, Sihao and Huang, Tiansheng and Liu, Gaowen and Kompella, Ramana Rao and Ilhan, Fatih and Tekin, Selim Furkan and Xu, Yichang and Yahn, Zachary and Liu, Ling},
  journal={arXiv preprint arXiv:2404.02039},
  year={2024}
}

@article{xu2024gameagentsurvey,
  title={A survey on game playing agents and large models: Methods, applications, and challenges},
  author={Xu, Xinrun and Wang, Yuxin and Xu, Chaoyi and Ding, Ziluo and Jiang, Jiechuan and Ding, Zhiming and Karlsson, B{\"o}rje F},
  journal={arXiv preprint arXiv:2403.10249},
  year={2024}
}

@article{dvcsurvey,
  title={Dense video captioning: A survey of techniques, datasets and evaluation protocols},
  author={Qasim, Iqra and Horsch, Alexander and Prasad, Dilip},
  journal={ACM Computing Surveys},
  volume={57},
  number={6},
  pages={1--36},
  year={2025},
  publisher={ACM New York, NY}
}

@article{kuhn1955hungarian,
  title={The Hungarian method for the assignment problem},
  author={Kuhn, Harold W},
  journal={Naval research logistics quarterly},
  volume={2},
  number={1-2},
  pages={83--97},
  year={1955},
  publisher={Wiley Online Library}
}

@article{carion2025sam,
  title={Sam 3: Segment anything with concepts},
  author={Carion, Nicolas and Gustafson, Laura and Hu, Yuan-Ting and Debnath, Shoubhik and Hu, Ronghang and Suris, Didac and Ryali, Chaitanya and Alwala, Kalyan Vasudev and Khedr, Haitham and Huang, Andrew and others},
  journal={arXiv preprint arXiv:2511.16719},
  year={2025}
}

@article{hurst2024gpt4o,
  title={Gpt-4o system card},
  author={Hurst, Aaron and Lerer, Adam and Goucher, Adam P and Perelman, Adam and Ramesh, Aditya and Clark, Aidan and Ostrow, AJ and Welihinda, Akila and Hayes, Alan and Radford, Alec and others},
  journal={arXiv preprint arXiv:2410.21276},
  year={2024}
}

@misc{praw_docs,
  author = {Bryce Boe},
  title = {PRAW: The Python Reddit API Wrapper Documentation},
  year = {2023},
  howpublished =  {\url{https://praw.readthedocs.io/}},
}

@book{schutze2008f1,
  title={Introduction to information retrieval},
  author={Sch{\"u}tze, Hinrich and Manning, Christopher D and Raghavan, Prabhakar},
  volume={39},
  year={2008},
  publisher={Cambridge University Press Cambridge}
}

@misc{gemini2,
  author = {Google},
  title = {Introducing Gemini 2.0: our new AI model for the agentic era},
  howpublished = {\url{https://blog.google/innovation-and-ai/models-and-research/google-deepmind/google-gemini-ai-update-december-2024/}},
  year = {2024},
}

@misc{claude3.5,
  author = {Anthropic},
  title = {Model Card Addendum: Claude 3.5 Haiku and Upgraded Claude 3.5 Sonnet},
  year = {2024},
  howpublished = {\url{https://assets.anthropic.com/m/1cd9d098ac3e6467/original/Claude-3-Model-Card-October-Addendum.pdf}},
}

@misc{nova,
  author = {Amazon Artificial General Intelligence},
  title = {The Amazon Nova Family of Models: Technical Report and Model Card},
  year = {2024},
  howpublished = {\url{https://www.amazon.science/publications/the-amazon-nova-family-of-models-technical-report-and-model-card}}
}

@article{Qwen2.5-VL,
  title={Qwen2.5-VL Technical Report},
  author={Bai, Shuai and Chen, Keqin and Liu, Xuejing and Wang, Jialin and Ge, Wenbin and Song, Sibo and Dang, Kai and Wang, Peng and Wang, Shijie and Tang, Jun and Zhong, Humen and Zhu, Yuanzhi and Yang, Mingkun and Li, Zhaohai and Wan, Jianqiang and Wang, Pengfei and Ding, Wei and Fu, Zheren and Xu, Yiheng and Ye, Jiabo and Zhang, Xi and Xie, Tianbao and Cheng, Zesen and Zhang, Hang and Yang, Zhibo and Xu, Haiyang and Lin, Junyang},
  journal={arXiv preprint arXiv:2502.13923},
  year={2025}
}

@article{intern2.5vl,
  title={Expanding Performance Boundaries of Open-Source Multimodal Models with Model, Data, and Test-Time Scaling},
  author={Chen, Zhe and Wang, Weiyun and Cao, Yue and Liu, Yangzhou and Gao, Zhangwei and Cui, Erfei and Zhu, Jinguo and Ye, Shenglong and Tian, Hao and Liu, Zhaoyang and others},
  journal={arXiv preprint arXiv:2412.05271},
  year={2024}
}

@misc{ui-tars-15-seed,
  title = {UI-TARS-1.5},
  author = {ByteDance Seed},
  howpublished = {\url{https://seed-tars.com/1.5}},
  year = {2025},
}

@article{llavaonevision,
  	title={LLaVA-OneVision: Easy Visual Task Transfer},
  	author={Li, Bo and Zhang, Yuanhan and Guo, Dong and Zhang, Renrui and Li, Feng and Zhang, Hao and Zhang, Kaichen and Li, Yanwei and Liu, Ziwei and Li, Chunyuan},
  	journal={arXiv preprint arXiv:2408.03326},
  	year={2024}
}

\appendix
\section{Details on \dataname{}}
\subsection{Statistics}
\label{appendix:stats}
\begin{figure}[ht]
  \centering
  \includegraphics[width=0.75\columnwidth]{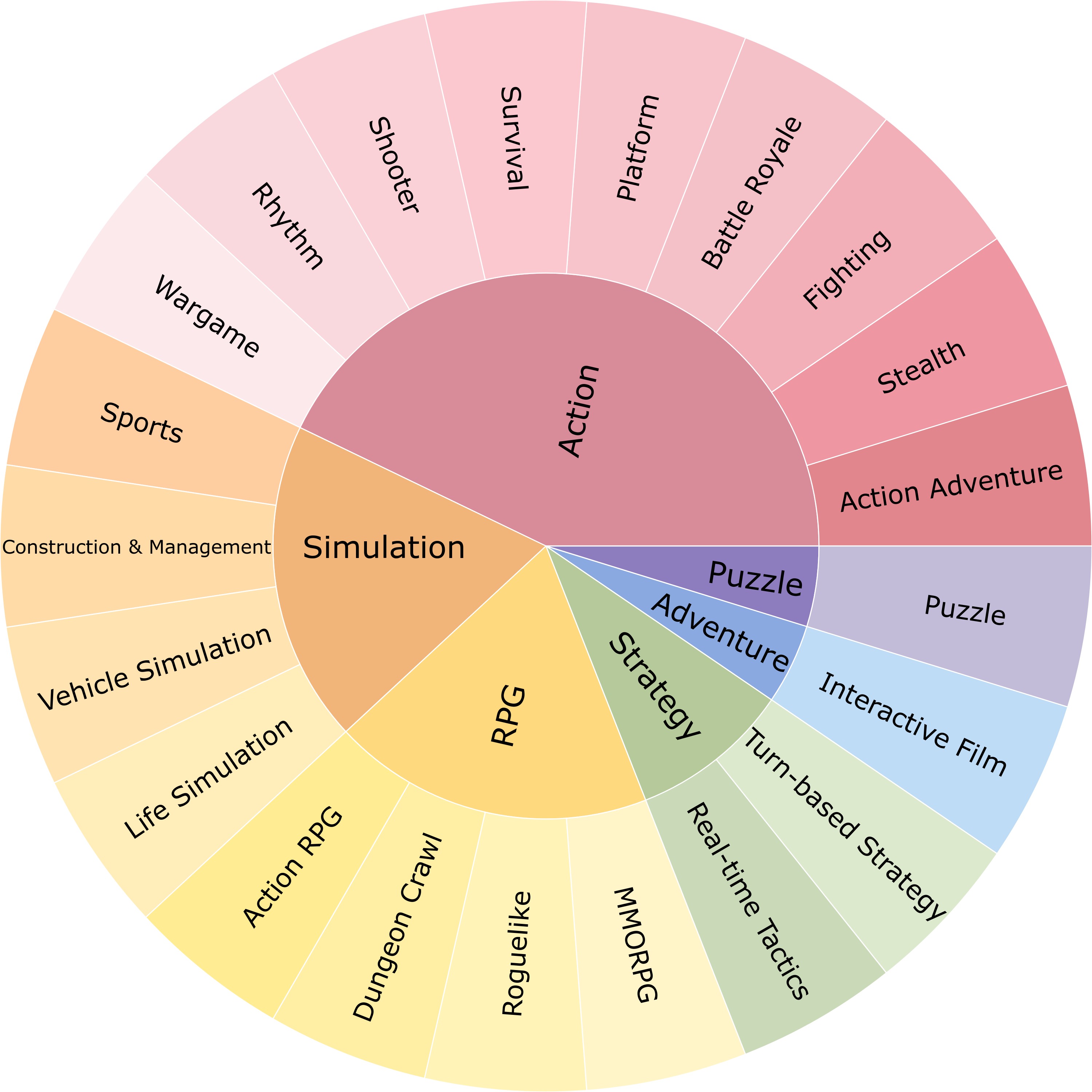}
  \caption{Taxonomy of \dataname.
  }
  \label{fig:data-genre}
  \Description{A pie chart illustrating the percentage breakdown of different video game genres included in the \dataname{} dataset, such as Action, Adventure, and RPG.}
\end{figure}

To support a diverse and balanced benchmark, we organize the selected games in \dataname{} into a taxonomy of genres and subgenres, as shown in Figure~\ref{fig:data-genre}. Specifically, the benchmark covers six major genres: \textit{Action}, \textit{Simulation}, \textit{RPG}, \textit{Strategy}, \textit{Puzzle}, and \textit{Adventure}. This taxonomy is used to guide genre-aware sampling, which helps reduce over-representation from a small number of dominant game types while improving coverage of underrepresented categories.

Table~\ref{tab:genre_subgenre_dist} further shows the video distribution across genres and subgenres. Action games account for the largest portion of the benchmark, covering 3,834 videos (73.20\%), with \textit{Action Adventure} and \textit{Shooter} as the two largest subgenres. Simulation is the second largest genre with 974 videos (18.59\%), followed by RPG with 313 videos (5.98\%). Smaller but still meaningful portions come from Strategy, Puzzle, and Adventure. This long-tail distribution reflects the natural composition of community-reported gameplay glitches, while our sampling strategy still preserves diversity across different gameplay styles and game mechanics.

\begin{table}[ht]
  \centering
  \caption{Video distribution by genre and subgenre.}
  \label{tab:genre_subgenre_dist}
  \setlength{\tabcolsep}{6pt}
  \begin{tabular}{llrr}
    \toprule
    Genre & Subgenre & Videos & Share (\%) \\
    \midrule

    \rowcolor{gray!15}
    \textbf{Action} &  & \textbf{3834} & \textbf{73.20} \\
     & Action Adventure & 1533 & 29.27 \\
     & Shooter & 1368 & 26.12 \\
     & Survival & 264 & 5.04 \\
     & Fighting & 260 & 4.96 \\
     & Battle Royale & 163 & 3.11 \\
     & Stealth & 133 & 2.54 \\
     & Platform & 71 & 1.36 \\
     & Wargame & 30 & 0.57 \\
     & Rhythm & 12 & 0.23 \\
    \addlinespace

    \rowcolor{gray!15}
    \textbf{Simulation} &  & \textbf{974} & \textbf{18.59} \\
     & Vehicle Simulation & 510 & 9.74 \\
     & \begin{tabular}[c]{@{}l@{}}Construction and \\ Management\end{tabular} & 258 & 4.93 \\
     & Sports & 140 & 2.67 \\
     & Life Simulation & 66 & 1.26 \\
    \addlinespace

    \rowcolor{gray!15}
    \textbf{RPG} &  & \textbf{313} & \textbf{5.98} \\
     & Action RPG & 264 & 5.04 \\
     & Dungeon Crawl & 20 & 0.38 \\
     & MMORPG & 19 & 0.36 \\
     & Roguelike & 10 & 0.19 \\
    \addlinespace

    \rowcolor{gray!15}
    \textbf{Strategy} &  & \textbf{60} & \textbf{1.15} \\
     & Real-time Tactics (RTT) & 30 & 0.57 \\
     & Turn-based Strategy & 30 & 0.57 \\
    \addlinespace

    \rowcolor{gray!15}
    \textbf{Puzzle} &  & \textbf{47} & \textbf{0.90} \\
     & Puzzle & 47 & 0.90 \\
    \addlinespace

    \rowcolor{gray!15}
    \textbf{Adventure} &  & \textbf{10} & \textbf{0.19} \\
     & Interactive Film & 10 & 0.19 \\

    \bottomrule
  \end{tabular}
\end{table}

\begin{figure}[ht]
    \centering
    \includegraphics[width=0.85\columnwidth]{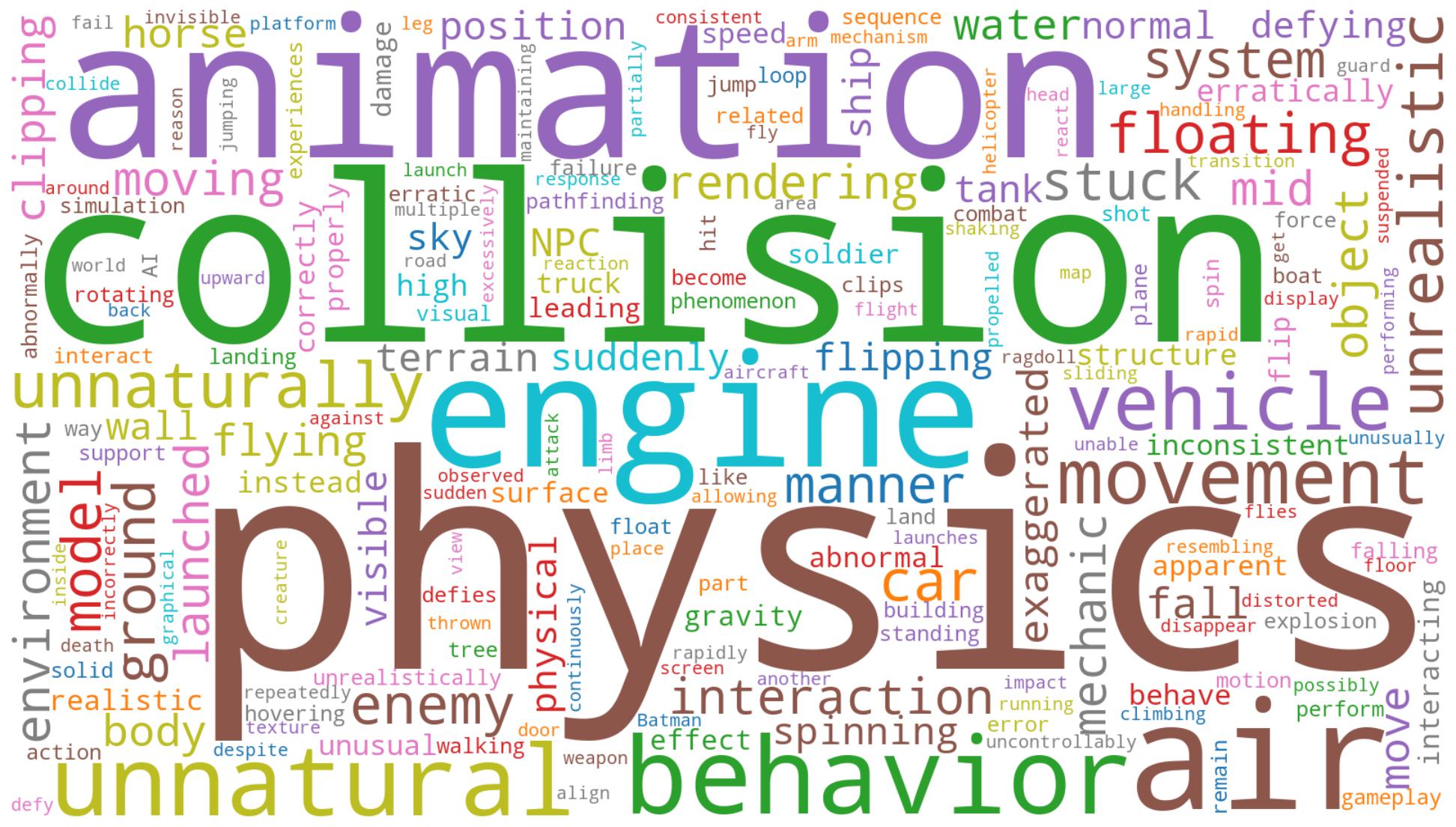}
    \caption{Word cloud of all the bug descriptions.}
    \label{fig:wordcloud}
\end{figure}

We also analyze the textual distribution of the annotated glitch descriptions. Figure~\ref{fig:wordcloud} presents a word cloud over all bug descriptions after removing stopwords. Frequent terms such as \textit{physics}, \textit{collision}, \textit{animation}, \textit{engine}, \textit{vehicle}, and \textit{character} suggest that \dataname{} covers a broad range of glitch types, including physics failures, rendering issues, animation errors, and unexpected object interactions. This observation is consistent with the goal of the benchmark: evaluating whether a model can understand and describe diverse glitch phenomena in open-ended gameplay videos.

\subsection{More on Annotation}
\label{appendix:anno}

\begin{table*}[ht]
\centering
\caption{Tool library for our agentic framework.}
\begin{tabular}{lp{14cm}}
\toprule
\textbf{Tool Name} & \textbf{Description} \\
\midrule
\texttt{vqa} & Submit the full stitched window image with a text question to the VLM and return a free-form answer. \\
\addlinespace
\texttt{zoom\_in} & Crop a specified region from one or more raw frames, upscale the result, and chain a VQA call on the cropped output. Used when the anomaly occupies a small portion of the full image. \\
\addlinespace
\texttt{object\_tracking} & Track a specified object across frames using SAM3, then automatically run physics analysis on the resulting centroid sequence. Physics analysis computes per-frame velocity and acceleration, and applies four rule-based anomaly detectors: (1)~\textit{position\_jump}: inter-frame displacement $> 20\%$ of the frame diagonal; (2)~\textit{velocity\_spike}: inter-frame speed change $> 500$\,px/s; (3)~\textit{motion\_freeze}: displacement $< 1$\,px for $\geq 4$ consecutive frames; (4)~\textit{jittering}: direction reversals in $\geq 50\%$ of consecutive frame-pair steps. \\
\bottomrule
\end{tabular}
\label{tab:tools}
\end{table*}

\begin{figure*}[ht]
\centering
\includegraphics[width=0.97\linewidth]{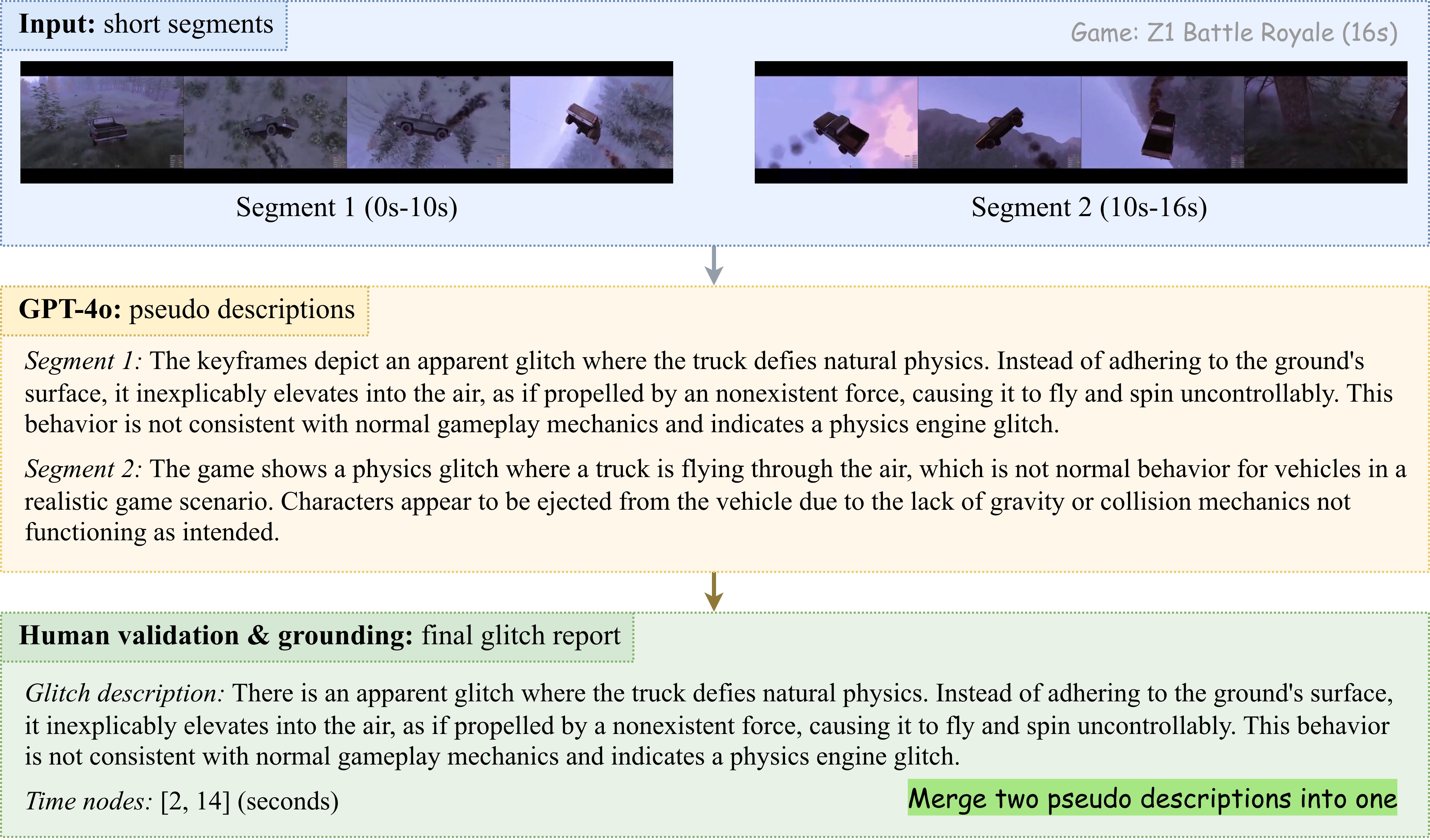}
\caption{An example of the annotation process.}
\label{fig:anno_example}
\end{figure*}

\textbf{Example of annotation process.}
Figure~\ref{fig:anno_example} shows a concrete example of our annotation process. The input video is first divided into short segments (< 10 seconds), and GPT-4o generates pseudo glitch descriptions independently for each segment. In this example, both segments describe a similar phenomenon where a truck exhibits physically implausible flying behavior. During human validation, annotators recognize that these segment-level descriptions correspond to the same underlying glitch. They therefore merge them into a single video-level glitch report, refine the description for clarity and correctness, and assign unified temporal boundaries that span the full duration of the event. This example illustrates how fragmented segment-level predictions are consolidated into coherent, temporally grounded annotations in \dataname{}.

\begin{figure*}[ht]
\centering
\includegraphics[width=0.97\linewidth]{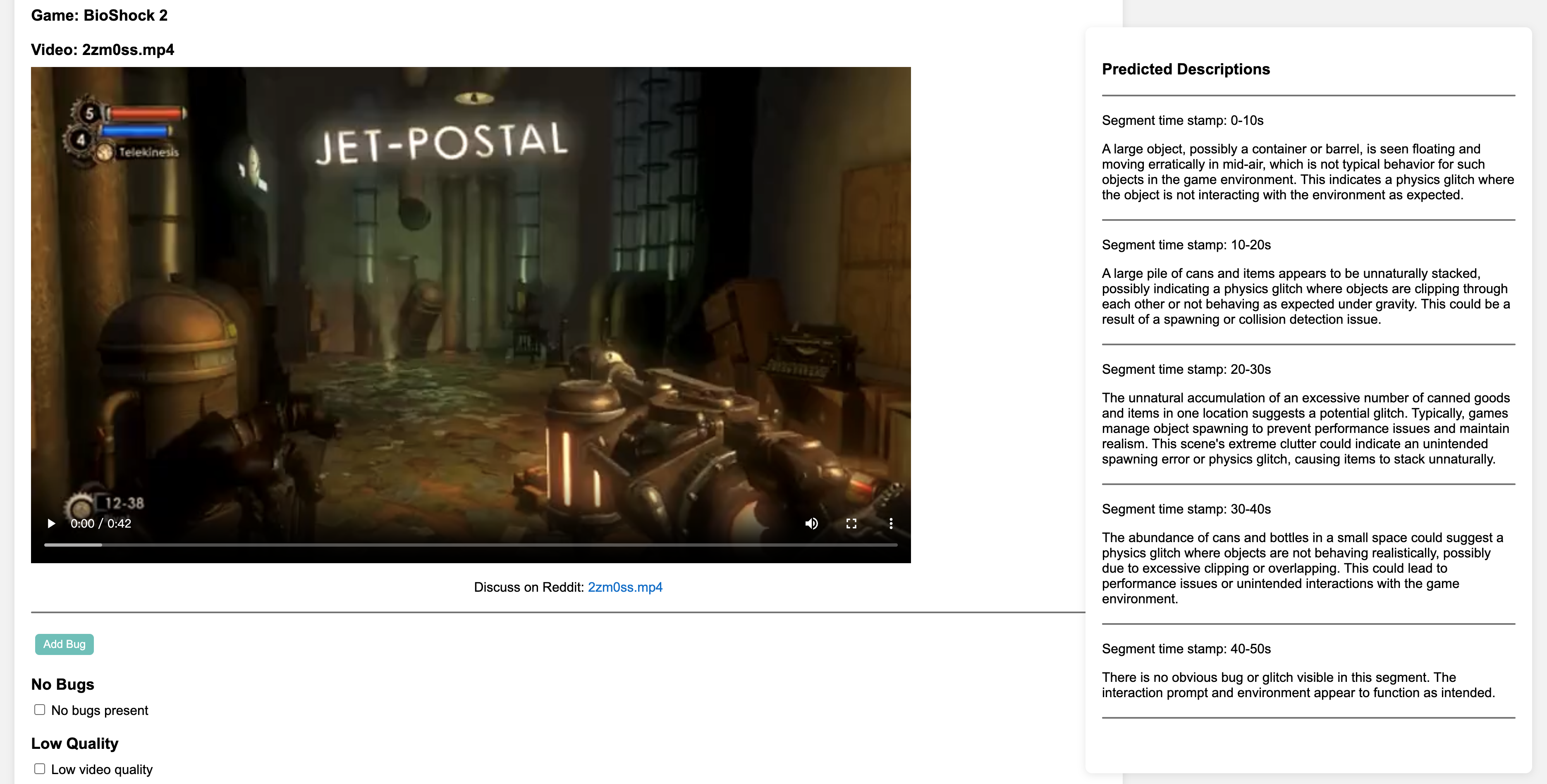}
\caption{Annotation interface used for \dataname{}.}
\label{fig:anno_interface}
\end{figure*}

\textbf{Annotation interface.}
To support efficient and consistent annotation, we develop a multi-user annotation interface, as illustrated in Figure~\ref{fig:anno_interface}. The interface presents the gameplay video together with model-generated segment-level pseudo descriptions. For each video, annotators can review the playback, inspect segment-wise predictions with corresponding timestamps, and directly edit descriptions or assign temporal boundaries. The interface also provides simple controls for adding, merging, or discarding glitch annotations, enabling streamlined video-level labeling.

\textbf{Annotator expertise.}
All annotations are conducted by volunteers with prior experience in gaming and familiarity with common glitch phenomena, such as physics violations, rendering artifacts, and abnormal object interactions. While the annotators are not professional QA engineers, they possess sufficient domain knowledge to reliably identify and describe glitches in gameplay videos. To ensure annotation quality, all samples undergo careful review and refinement during the validation process.

\section{Experiment Details}
\subsection{Full Experimental Setup}
\label{appendix:detailed_setup}
\textbf{Video preprocessing.}
Input videos are processed using time-based frame sampling via OpenCV, extracting frames at a fixed rate of 4 FPS (i.e., one frame every 0.25
seconds), regardless of the original video frame rate. This ensures uniform temporal coverage across videos with varying native frame rates. Extracted frames are organized into non-overlapping windows of 8 frames each, yielding a temporal span of 2 seconds per window. Within each window, the 8 frames are stitched into a single composite image arranged in a 2-row $\times$ 4-column grid layout, with each frame labeled by its sequential index in the top-left corner. Stitched
images are saved as JPEG. This multi-frame
stitching design allows the model to observe temporal progression within a single model call.
\par \textbf{Rough detection.}
The Scanner is configured with a sampling temperature of 0.5 and a maximum output length of 512 tokens. In addition to the glitch label and
confidence score, the Scanner produces a \texttt{game\_context} field --- a 2–4 sentence
natural language description of the game genre, environment, and visible mechanics, which is propagated to all downstream modules as a lightweight retrieval-augmented context.
\par \textbf{Fine-grained verification.}
This module operates on each Scanner-flagged window through an iterative Planner $\rightarrow$ Executor $\rightarrow$ Reflector loop, with a maximum of 5 iterations per window. The loop terminates early if the Judge's \texttt{final\_confidence} reaches the threshold of 0.80. All sub-agents (Planner, Advocate, Skeptic, Judge) share a single MLLM client configured with temperature 0.5 and max tokens 1,024.
\par \textbf{Temporal grounding.}
The Grounder consolidates per-window outputs from the previous module into temporally contiguous glitch records through two steps. \textbf{1)} Similarity clustering. Confirmed glitch windows are processed in chronological order. Each new glitch description is compared against all
existing cluster descriptions via an MLLM call; if the model returns \texttt{yes}, the window is merged into that cluster, otherwise a new
cluster is created (greedy first-match assignment). Similarity is judged on three axes: entity appearance similarity, anomaly type similarity, and behavioral similarity. \textbf{2)} Bidirectional propagation. For each confirmed glitch cluster, the Grounder expands the cluster both forward and backward by querying the model with the glitch description and the adjacent window's image. A window is added to the cluster if the model returns \texttt{yes}; propagation stops at the first non-matching window.
\par \textbf{Final report generation.}
The Summarizer receives the list of \texttt{GlitchRecord} objects from the Grounder and generates the final structured report. Each record's frame-level occurrences are converted to second-level timestamps using the extraction FPS
of 4.0. The Summarizer model is configured with temperature 0.5 and max tokens 512. The output report is a JSON file containing the video name, game name, a boolean \texttt{no\_bugs} flag, a list of natural-language bug descriptions, and a list of \texttt{[start\_sec, end\_sec]} time intervals corresponding to each bug.

\subsection{Tool Implementation}
\label{appendix:tools}
Table~\ref{tab:tools} demonstrates our tool library. Our agentic framework is equipped with three investigation tools --- \texttt{vqa}, \texttt{zoom\_in}, and \texttt{object\_tracking}. For \texttt{object\_tracking}, the SAM3 tracker is lazily initialized on first use and its session is shared across all windows in a batch; it is allocated to a separate GPU from the main VLM.

\subsection{Prompts for \methodname}
\label{appendix:prompts}
We provide the core prompts used in our framework, including those for the Scanner, Planner, Reflector (Advocate, Skeptic, Judge), Grounder, and Summarizer.

\begin{PromptBox}{Scanner (Rough Detection) Prompt}
\textbf{Role.} You are a video game QA expert performing initial glitch screening.

\textbf{Input.} A stitched image of $N$ consecutive frames, each labelled \texttt{\#0, \#1, \ldots} in temporal order.

\textbf{Task.} Read frames in sequence, track how objects and characters change across frames, and determine whether a glitch is present. Flag uncertain cases with low confidence rather than dismissing them---false negatives are more costly than false positives.

\textbf{Glitch categories:}
\begin{itemize}[leftmargin=*]
  \item \textit{Visual}: texture/lighting errors, model distortion, flickering, camera clipping.
  \item \textit{Physics}: clipping through surfaces, abnormal launch or flight, floating, jittering.
  \item \textit{Game Logic}: AI stuck, animation frozen while character moves, interaction failures.
  \item \textit{Other}: network desync, projectile anomalies, unclassifiable behaviors.
\end{itemize}

\textbf{Confidence scale.} 0.85--1.0 obvious; 0.65--0.85 clear with minor ambiguity; 0.45--0.65 suspicious; 0.25--0.45 ambiguous; 0.0--0.25 likely normal. Prefer flagging suspicious cases at 0.5--0.7 over dismissing them.

\textbf{\texttt{game\_context} field (required in every response).} Describe in 2--4 sentences the game genre, environment/setting, key visible elements, and observed mechanics. This field serves as a knowledge base for later stages.

\textbf{Output format:}
\begin{lstlisting}[
    basicstyle=\ttfamily\small,
    breaklines=true,
    columns=flexible, 
    keepspaces=true,
    showstringspaces=false
]
{
  "has_glitch": true,
  "category": "Physics|Visual|Game Logic|Other",
  "visual_cues": "Red car at constant height in frames #4-7, no support",
  "reasoning":   "Floating without ground contact, violates gravity",
  "frame_range": [4, 5, 6, 7],
  "confidence":  0.87,
  "game_context": "Open-world racing game. Urban road with traffic. ..."
}
// If no glitch: omit category/visual\_cues/frame\_range; set has\_glitch to false.
\end{lstlisting}
\end{PromptBox}
\label{fig:prompt_scanner}

\begin{PromptBox}{Planner Prompt}
\textbf{Role.} You are the Planner in a video game glitch detection system.

\textbf{Input.} The Scanner's glitch hypothesis (\texttt{category}, \texttt{visual\_cues}, \texttt{confidence}, \texttt{game\_context}) and a history of prior tool calls with their results.

\textbf{Task.} Select the next investigation tool and formulate a specific, neutral question. \textbf{Always call \texttt{vqa} on the first iteration.}

\textbf{Available tools:}
\begin{itemize}[leftmargin=*]
  \item \texttt{vqa}: Ask a visual question about the full stitched window image.
  \item \texttt{zoom\_in}: Crop and magnify a region of one or more frames, then ask a question. Use when full-frame VQA was vague or the glitch is spatially localized. Parameters: \texttt{frame\_index} (int or list), \texttt{region} (spatial name or \texttt{[x1,y1,x2,y2]}), \texttt{question}.
  \item \texttt{object\_tracking}: Track an object using SAM3; returns frame-by-frame bounding boxes, speed, and anomaly flags (position jump, motion freeze, velocity spike, jitter). Use for Physics glitches when VQA evidence is ambiguous. Parameter: \texttt{object\_description}---a short visual description of the object (e.g., ``red sports car''), \textbf{not} a description of the glitch.
\end{itemize}

\textbf{Question style.} Ask neutral, descriptive questions---not leading confirmations.
\begin{itemize}[leftmargin=*]
  \item Wrong: ``Is the car floating?''
  \item Right: ``Describe the car's vertical position relative to the ground.''
\end{itemize}
The \textbf{first question} must establish scene context: describe all visible objects, their appearance (color, shape, size), and what they are doing.

\textbf{Strategy.} Iteration~1: always \texttt{vqa}. Iteration~2: prefer \texttt{object\_tracking} for Physics, \texttt{zoom\_in} for Visual/Animation. Iteration~3+: conclude unless the Judge ruled \texttt{needs\_more\_evidence} with a specific follow-up question. Never repeat the same question.

\textbf{Output format:}
\begin{lstlisting}[
    basicstyle=\ttfamily\small,
    breaklines=true,
    columns=flexible, 
    keepspaces=true,
    showstringspaces=false
]
{ "reasoning": "...", "tool": "vqa", "question": "..." }
{ "reasoning": "...", "tool": "zoom_in", "frame_index": 5, "region": "bottom_center", "question": "..." }
{ "reasoning": "...", "tool": "object_tracking", "object_description": "red sports car" }
{ "reasoning": "...", "tool": "none", "conclusion": "ready_to_conclude" }
\end{lstlisting}
\end{PromptBox}
\label{fig:scanner_prompt}

\begin{PromptBox}{Reflector -- Advocate Prompt}
\textbf{Role.} You are the Advocate (prosecutor) in a video game glitch detection system.

\textbf{Input.} Scanner's glitch hypothesis and the current tool result (VQA answer, zoom result, or object tracking data).

\textbf{Task.} Build a case for why the observed behavior is a glitch.
\begin{enumerate}[leftmargin=*]
  \item Identify abnormal behaviors \textbf{directly visible in the tool result.}
  \item Explain why normal game mechanics cannot account for them.
  \item Cite the specific physics, visual, or logic being violated.
  \item Always include the \textbf{specific visual appearance} of the affected object (color, shape, type) so it can be distinguished from other objects.
\end{enumerate}

\textbf{Output format:}
\begin{lstlisting}[
    basicstyle=\ttfamily\small,
    breaklines=true,
    columns=flexible, 
    keepspaces=true,
    showstringspaces=false
]
{
  "supporting_evidence": ["evidence 1 with object appearance", "..."],
  "affected_object_appearance": "Red sports car with black roof",
  "argument":   "This is a glitch because...",
  "violated_rules": ["gravity", "collision"],
  "confidence_for_glitch": 0.85
}
\end{lstlisting}
\end{PromptBox}

\begin{PromptBox}{Reflector -- Skeptic Prompt}
\textbf{Role.} You are the Skeptic (defense) in a video game glitch detection system.

\textbf{Input.} The Scanner's glitch hypothesis and the current tool result (VQA answer, zoom result, or object tracking data).

\textbf{Task.} Build a case for why the observed behavior is intentional game design.
\begin{enumerate}[leftmargin=*]
  \item Propose alternative explanations (game mechanics, special abilities, physics features, art style choices).
  \item Identify missing context that could explain the behavior.
  \item Point out ambiguities in the evidence.
  \item \textbf{Challenge visual grounding:} ask whether the anomaly is concretely visible in the tool result, or whether the Advocate is repeating the Scanner's initial hypothesis without direct visual confirmation from the investigation.
\end{enumerate}

\textbf{Output format:}
\begin{lstlisting}[
    basicstyle=\ttfamily\small,
    breaklines=true,
    columns=flexible, 
    keepspaces=true,
    showstringspaces=false
]
{
  "alternative_explanations": ["explanation 1", "explanation 2"],
  "argument": "This could be normal because...",
  "missing_context": ["need to check X"],
  "confidence_for_normal": 0.45
}
\end{lstlisting}
\end{PromptBox}
\label{fig:scanner_prompt}

\begin{PromptBox}{Reflector -- Judge Prompt}
\textbf{Role.} You are the Judge (neutral arbiter) in a video game glitch detection system.

\textbf{Input.} Advocate output, Skeptic output, and the full tool result.

\textbf{Task.}
\begin{enumerate}[leftmargin=*]
  \item \textbf{Visual grounding check (do this first).} Verify that the alleged anomaly is directly observable in the tool results. If the Advocate's argument rests on the Scanner hypothesis rather than concrete investigation evidence, treat it as unsubstantiated.
  \item Evaluate both sides and rule: \textbf{glitch} (anomaly directly visible; Advocate provides specific evidence; Skeptic's explanations are implausible); \textbf{normal} (Skeptic provides a plausible explanation).
  \item Correct the glitch category if evidence points to a different type.
\end{enumerate}

\textbf{Confidence scale.} 0.85--1.0 overwhelming; 0.70--0.85 strong; 0.55--0.70 leaning with uncertainty; below 0.55 genuinely ambiguous.

\textbf{Output format:}
\begin{lstlisting}[
    basicstyle=\ttfamily\small,
    breaklines=true,
    columns=flexible, 
    keepspaces=true,
    showstringspaces=false
]
{
  "advocate_summary": "...",  "skeptic_summary": "...",
  "ruling": "glitch | normal",
  "reasoning": "The evidence favors X because...",
  "category": "Physics|Visual|Game Logic|Other",
  "category_corrected": false,  "correction_reason": "(if corrected)",
  "subtype": "Floating|Clipping|...",
  "final_confidence": 0.78,  "should_continue": false,
  "next_questions": ["question 1"]
}
// When ruling="glitch" and should_continue=false, also include:
{
  "description": "Object appearance + what the glitch looks like + scene location.",
  "supporting_evidence": ["..."],  "rejected_explanations": ["..."]
}
\end{lstlisting}
\end{PromptBox}

\begin{PromptBox}{Grounder -- Glitch Clustering Prompt}
\textbf{Task.} Determine whether the given anomaly description belongs to an existing glitch cluster.

\textbf{Criteria for similarity:} (1) similar entities---allow for minor observation variance; (2) similar anomaly type (physical, visual, animation); (3) similar abnormal behavior (e.g., float/flying, clipping/collision).

\textbf{Input.} \texttt{anomaly\_description} (new glitch) $+$ \texttt{existing\_descriptions} (cluster, in time order).

\textbf{Output format:}
\begin{lstlisting}[
    basicstyle=\ttfamily\small,
    breaklines=true,
    columns=flexible, 
    keepspaces=true,
    showstringspaces=false
]
{ "reasoning": "step-by-step similarity analysis", "judgement": "yes | no" }
\end{lstlisting}
\end{PromptBox}

\begin{PromptBox}{Grounder -- Bidirectional Propagation Prompt}
\textbf{Task.} Detect whether the following anomaly (or a similar one) is visible in the provided window image. Used to expand the temporal boundary of a confirmed glitch.

\textbf{Criteria for similarity:} (1) similar entities ((objects, characters, creatures)) --- might differ slightly due to observation variance; (2) similar anomaly type (physical, visual, animation).

\textbf{Input.} \texttt{anomaly\_description} (confirmed glitch) $+$ stitched window image.

\textbf{Output format:}
\begin{lstlisting}[
    basicstyle=\ttfamily\small,
    breaklines=true,
    columns=flexible, 
    keepspaces=true,
    showstringspaces=false
]
{ "reasoning": "step-by-step visual analysis of the image", "judgement": "yes | no" }
\end{lstlisting}
\end{PromptBox}

\begin{PromptBox}{Summarizer -- Final Report Generation Prompt}
\textbf{Role.} You are a video game glitch analyst.

\textbf{Input.}
\begin{lstlisting}[
    basicstyle=\ttfamily\small,
    breaklines=true,
    columns=flexible, 
    keepspaces=true,
    showstringspaces=false
]
Descriptions (chronological, same glitch): {descriptions}
Category: {category} | Subtype: {subtype} | Time range: {time_range}
\end{lstlisting}

\textbf{Task.} Summarize the fragmented descriptions into a single, clear, coherent description for the final report.
\begin{enumerate}[leftmargin=*]
  \item Identify the core glitch phenomenon across all descriptions.
  \item Write one coherent paragraph (2--4 sentences) that describes the visual/behavioral anomaly, mentions the \textbf{specific appearance} of the affected object, and states where in the scene it occurs.
  \item \textbf{Do not include} frame numbers, timestamps, JSON, or code blocks.
  \item Use natural, descriptive language.
\end{enumerate}

\textbf{Output.} The summarized description only, without any additional text.
\end{PromptBox}

\subsection{Example Execution Trace}
To better illustrate how \methodname{} operates in practice, we present a detailed execution trace on a representative example with multiple glitches. This case involves both collision-related errors (clipping through solid structures) and physics anomalies (launching and disappearance), making it a challenging scenario for reliable detection and grounding.
\par The example highlights the full pipeline of \methodname{}, including high-recall candidate screening, iterative verification with tool-assisted reasoning, and temporal grounding across disjoint segments. In particular, it demonstrates how the framework rejects false positives through multi-step reasoning (e.g., Window 0), while leveraging tools such as VQA and object tracking to resolve harder cases. The Grounder further consolidates fragmented detections into coherent glitch reports with accurate temporal spans.

\begin{TraceBox}{Execution Trace Example of \methodname{}}
\textbf{Game:} Steep \quad
\textbf{Video:} 5dq755.mp4

\textbf{Ground Truth:}

- Clipping through wooden structures (collision failure). Timestamps: [0, 5], [12, 15]

- Character deformation / teleportation. Timestamps: [5, 6], [8, 15]

\textbf{\textcolor{blue}{1. Rough Detection}}

[W0] GLITCH (Visual, 0.75) -- airborne anomaly  

[W1] GLITCH (Physics, 0.78) -- clipping through barrier  

[W2] GLITCH (Physics, 0.75) -- abnormal motion  

[W3] GLITCH (Physics, 0.75) -- unrealistic interaction 

[W4] GLITCH (Physics, 0.75) -- falling through terrain

[W5-7] No glitch

$\rightarrow$ 5 candidate glitch windows detected, covering both potential clipping and motion anomalies.

\textbf{Game-aware context}: Action-adventure game with snowy mountain terrain. The player character, dressed in a blue outfit, seems to be exploring or interacting with the environment, as indicated by the presence of an exploration camera icon.

\textbf{\textcolor{blue}{2. Fine-grained Verification}}

\textbf{Window 0}

Iteration 1:

- \textit{Planner} $\rightarrow$ VQA: "Describe ground interaction"

- Answer: Character appears sliding/falling with continuous ground contact

\textit{Advocate}: glitch (possible hovering / missing collision) 

\textit{Skeptic}: normal (intentional sliding or terrain interaction)  

\textit{Judge}: \textbf{NORMAL} (confidence = 0.65, below threshold)

Iteration 2:

- \textit{Planner} $\rightarrow$ zoom-in (ground contact region)

- Observation: no clear rendering or contact anomaly

$\rightarrow$ After multiple iterations (max=5), no strong evidence of rule violation 

\textbf{Final decision: NORMAL (false positive rejected)}

\textbf{Window 1--2}

- \textit{Planner} $\rightarrow$ VQA verification

- Evidence: character penetrates wooden barrier without resistance $\rightarrow$ Violates collision constraint.  

\textbf{Confirmed clipping glitch}

\textbf{Window 3}

- \textit{Planner} $\rightarrow$ VQA verification

- Evidence: motion inconsistent with expected gravity-driven trajectory $\rightarrow$ Indicates unstable physics behavior.  

\textbf{Confirmed physics glitch}

\textbf{Window 4}

- \textit{Planner} $\rightarrow$ object tracking

- Observation: tracked character abruptly disappears mid-trajectory $\rightarrow$ Strong evidence of physics engine failure (trajectory discontinuity).  

\textbf{Confirmed physics glitch}

\textbf{\textcolor{blue}{3. Temporal Localization}}

Initial clustering based on semantic similarity:

- Cluster 1: [W1, W2] (clipping-related)

- Cluster 2: [W3, W4] (teleportation-related)

Bidirectional propagation with visual consistency check:

- Cluster 1 $\rightarrow$ [W1, W2, W4, W5, W6]

- Cluster 2 $\rightarrow$ [W2, W3, W4, W5, W6, W7]

$\rightarrow$ Expands temporal coverage to capture long-range and recurring glitches.

\textbf{\textcolor{blue}{4. Final Glitch Report Generation}}

Bug 1 (Clipping):
Character phases through a wooden barrier, violating collision constraints.  
Timestamps: [2--5], [8--13]

Bug 2 (Teleportation):
Character launches and disappears without physical cause.  
Timestamps: [4--15]
\end{TraceBox}

\end{document}